\documentclass{article}
\usepackage{authblk}
\usepackage{amsmath}
\usepackage[ruled, vlined, linesnumbered]{algorithm2e}
\usepackage{float}
\usepackage{hyperref}
\usepackage{PRIMEarxiv}

\usepackage[utf8]{inputenc} 
\usepackage[T1]{fontenc}    
\usepackage{hyperref}       
\usepackage{url}            
\usepackage{booktabs}       
\usepackage{amsfonts}       
\usepackage{nicefrac}       
\usepackage{microtype}      
\usepackage{lipsum}
\usepackage{fancyhdr}       
\usepackage{graphicx}       
\graphicspath{{media/}}     

\title{Hedging with Sparse Reward Reinforcement Learning
}

\author[1,2]{Yiheng Ding}
\author[3]{Gangnan Yuan}
\author[1]{Dewei Zuo}
\author[1,2]{Ting Gao\thanks{Corresponding author. Email: \href{mailto:tgao0716@hust.edu.cn}{tgao0716@hust.edu.cn}} ~}

\affil[1]{School of Mathematics and Statistics, Huazhong University of Science and Technology, Wuhan 430074, China}
\affil[2]{Center for Mathematical Science, Huazhong University of Science and Technology, Wuhan 430074, China}
\affil[3]{Great Bay University, Dongguan 523000, China}

\begin{document}
\maketitle

\begin{abstract}
Derivatives, as a critical class of financial instruments, isolate and trade the price attributes of risk assets such as stocks, commodities, and indices, aiding risk management and enhancing market efficiency. However, traditional hedging models, constrained by assumptions such as continuous trading and zero transaction costs, fail to satisfy risk control requirements in complex and uncertain real-world markets.

With advances in computing technology and deep learning, data-driven trading strategies are becoming increasingly prevalent. This thesis proposes a derivatives hedging framework integrating deep learning and reinforcement learning. The framework comprises a probabilistic forecasting model and a hedging agent, enabling market probability prediction, derivative pricing, and hedging.

Specifically, we design a spatiotemporal attention-based probabilistic financial time series forecasting Transformer to address the scarcity of derivatives hedging data. A low-rank attention mechanism compresses high-dimensional assets into a low-dimensional latent space, capturing nonlinear asset relationships. The Transformer models sequential dependencies within this latent space, improving market probability forecasts and constructing an online training environment for downstream hedging tasks.

Additionally, we incorporate generalized geometric Brownian motion to develop a risk-neutral pricing approach for derivatives. We model derivatives hedging as a reinforcement learning problem with sparse rewards and propose a behavior cloning-based recurrent proximal policy optimization (BC-RPPO) algorithm. This pretraining-finetuning framework significantly enhances the hedging agent's performance. Numerical experiments in the U.S. and Chinese financial markets demonstrate our method's superiority over traditional approaches.

\end{abstract}


\section{Introduction}
\subsection{Research Background and Significance}
As financial markets continue to evolve, derivatives trading has become a vital component of risk management and portfolio optimization. Derivatives allow investors to hedge against market fluctuations by isolating and trading specific risk factors, such as volatility and interest rate changes. However, traditional hedging strategies, such as Delta hedging based on the Black-Scholes model \cite{black1973pricing}, rely on assumptions like continuous trading and frictionless markets, which are unrealistic in real-world financial environments.

With the advancement of deep learning and reinforcement learning, data-driven approaches have emerged as powerful alternatives to traditional models. These techniques enable market participants to adapt to changing market conditions dynamically. Reinforcement learning (RL) is particularly well-suited for derivatives hedging, as it formulates the problem as a sequential decision-making process where the agent learns optimal hedging strategies by interacting with a simulated environment.

This research aims to develop a data-driven derivatives hedging framework that leverages probabilistic forecasting models and deep reinforcement learning. By integrating market probability predictions with an RL-based hedging agent, the proposed framework seeks to improve hedging effectiveness and mitigate risk in complex financial markets.

\subsection{Literature Review}
In addition to its success in Atari games, RL has been increasingly applied in financial trading. However, training RL agents directly in real markets is costly, so most training is done using simulated environments based on historical data. For instance, Theate and Ernst \cite{theate2021application} used DQN to train a trading agent with artificially generated sample trajectories from historical data. Massahi and Mahootchi \cite{massahi2024deep} developed a market simulator based on historical order book data and enhanced it with LSTM and DQN for algorithmic trading in futures markets. Wu et al. \cite{wu2020adaptive} incorporated technical factors from historical data into the environment and combined DQN with DDPG to enhance performance in multiple major markets.

Derivatives play a vital role in risk management, offering high leverage and complex risk structures, making effective hedging essential. The primary objective of derivatives hedging is to reduce or eliminate market risk, thus protecting investors from adverse market fluctuations. The Black-Scholes model \cite{black1973pricing} was the first to provide a Delta-hedging approach for European options, which hedges option price changes by trading the underlying asset. Over the last two decades, data-driven hedging strategies have become mainstream.

Buehler et al. \cite{buehler2019deep} extended this framework using deep learning for derivatives hedging in the presence of trading frictions, providing a theoretical foundation for deep learning in hedging. Zhu and Diao \cite{zhu2023stochastic} further developed a GRU-based hedging framework under rough volatility, demonstrating its superiority over traditional deep learning techniques in non-Markov environments.

The integration of reinforcement learning into derivatives hedging started earlier. In 2017, Halperin \cite{halperin2017qlbs} applied Q-learning to combine RL with the Black-Scholes model for derivatives hedging. Subsequently, Halperin \cite{halperin2019qlbs} conducted numerical experiments on pricing and hedging within the Black-Scholes framework using dynamic programming, addressing issues like volatility smiles.

In value-based methods, Cao et al. \cite{cao2021deep} used Double Q-learning to separately track the mean and variance of hedging costs, developing a strategy superior to Delta hedging under trading costs. Later, Cao et al. \cite{cao2023gamma} combined DQN with quantile regression to hedge options with random maturities, managing portfolio risks like Gamma and Vega. Buehler \cite{buehler2022deep} demonstrated the existence of a finite solution to the Bellman equation in the framework of futures, forwards, and options hedging, showing its relationship with classical deep hedging methods.

In policy gradient methods, Xiao et al. \cite{xiao2021optimal} applied Monte Carlo policy gradient to hedge options, accounting for non-zero risk-free rates and trading costs. Du et al. \cite{du2020deep} used PPO in a discrete trading environment with nonlinear transaction costs, proving that PPO outperforms other methods in option hedging tasks.

In risk management for derivatives portfolios, Vittori et al. \cite{vittori2020option} used safe reinforcement learning to train agents with varying risk aversions, constructing an efficient frontier in volatility and profit-loss space for hedging different types of options. Wu and Jaimungal \cite{wu2023robust} studied the application of risk-averse reinforcement learning in path-dependent derivatives, demonstrating how optimal hedging strategies can change when agents switch from risk-averse to risk-seeking behavior.

Recent research also extends to American and exotic options. Pickard et al. \cite{pickard2024optimizing} used DDPG to hedge American put options, achieving better results than traditional methods in both simulated and market-calibrated stochastic volatility models. Chen et al. \cite{yuana2024hedging} explored RL-based hedging for barrier options, showing its superiority over traditional methods with trading costs in experimental settings.

\subsection{Existing Challenges in Derivatives Hedging}
Despite the progress in derivatives pricing and hedging, several challenges remain:

\begin{itemize}
    \item \textbf{Market Assumptions}: Classical models such as Black-Scholes \cite{black1973pricing} and Heston \cite{heston1993closed} assume constant volatility and continuous trading, which do not hold in practice. Market frictions, liquidity constraints, and stochastic volatility introduce significant deviations from these assumptions.
    \item \textbf{Data Scarcity}: High-quality derivatives hedging datasets are limited, making it difficult to train robust hedging strategies. While historical data can be used, market conditions are constantly changing, requiring adaptive models that generalize across different regimes.
    \item \textbf{Model Dependency}: Many existing reinforcement learning-based hedging strategies rely on theoretical pricing models to compute rewards, leading to biases if the assumed model deviates from market reality \cite{cao2021deep,sharma2024hedging}.
    \item \textbf{Exploration vs. Exploitation}: RL-based methods require extensive exploration to learn optimal hedging strategies. However, excessive exploration in real markets can be costly. Efficient training methods, such as pretraining on simulated data and fine-tuning with real data, are necessary to address this issue \cite{zhu2023stochastic}.
\end{itemize}

\section{Stock Return Prediction} 

\subsection{Deterministic prediction and probabilistic prediction}

Due to the dimensional differences in prices across different stocks, we often transform price data into logarithmic return data for use. Let $S_{i, t}$ denote the price of asset $i$ at time $t$, and we define the logarithmic return as:
\begin{equation}
r_{i, t} = \ln(S_{i, t+1}) - \ln(S_{i, t})
\label{eq:log_return}
\end{equation}
where $i \in \{1, 2, ..., N\}$ represents the index of the asset, and $t \in {1, 2, ..., T}$ represents the time steps. People commonly use candlestick charts to describe price fluctuations in financial markets, we transform the opening, highest, lowest, and closing price into logarithmic returns to be input into the model. Specifically, we define the combined data as: $r^\text{data}_t = \text{Concat} \left( r^O_{i,t}, r^H_{i,t}, r^L_{i,t}, r^C_{i,t} \right) \in \mathbb{R}^{N \times 4}$, where

\begin{equation}
\begin{aligned}
r^O_{i, t} &= \ln(O_{i, t+1}) - \ln(C_{i, t})  \\
r^H_{i, t} &= \ln(H_{i, t+1}) - \ln(C_{i, t})  \\
r^L_{i, t} &= \ln(L_{i, t+1}) - \ln(C_{i, t})  \\
r^C_{i, t} &= \ln(C_{i, t+1}) - \ln(C_{i, t})  
\end{aligned}
\label{eq:log_returns}
\end{equation}

$O, H, L, C$ represent the opening, highest, lowest, and closing prices respectively. 

Over the past four decades, a prevalent approach among quantitative funds has been to achieve excess returns by forecasting assets. For instance, explaining the disparities in stock returns through the momentum effect of stocks. We aid investment decisions by forecasting the future returns or prices of assets. They predict future returns to facilitate investment decision-making. Deterministic prediction is point prediction, typically involving a loss function,
\begin{equation}
Loss(\theta) = \mathbb{E} \Vert r^\text{data}_t - \hat{r}_t(\theta) \Vert ^ 2
\label{eq:deterministic_loss}
\end{equation}
where $\hat{r(\theta)}$ is the prediction generated using a parametrized model (e.g., linear model, tree model, neural networks, etc.). Deterministic predictions are usually most accurate for the next period, so a rolling method is used to recursively generate the prediction for period $t$ using data from periods 0 to $t-1$.
Probabilistic predictions aim to forecast the distribution of returns. Although in reality, we can only see one trajectory of prices, there are infinitely many possible movements for future prices. The second-order information (such as $\mathbb{E}(r_t^2)$) or higher-order information (like skewness and kurtosis) about stock volatility is contained in the distribution information. Typically, there is a loss function for this as well:

\begin{equation}
Loss(\theta) = - \sum_{t} \text{log}P_{\hat{Q}} (r^\text{data}_t;\theta)
\label{eq:probabilistic_loss}
\end{equation}
where $\hat{Q}$ represents the distribution of parameter $\theta$, and $P$ denotes the probability associated with the given data $r^\text{data}_t$. By assuming the form of the distribution beforehand and modeling the parameters of the distribution using linear or non-linear models, the training is done through maximum likelihood loss. By recursively sampling from the model, we can obtain any number of predicted sequences, representing the possible trajectories of asset returns over a future period.

Modeling the returns of risky assets as a normal distribution is appropriate and suitable due to the Central Limit Theorem \cite{kamalov2021financial}. We only need to model parameterized $\mu_{i, t}(\theta, h_{\cdot, 0:t-1})$ and $\sigma_{i, t}(\theta, h_{\cdot, 0:t-1})$, with:
\begin{equation}
r_{i, t} \sim \mathcal{N}(\mu_{i, t}(\theta, h_{\cdot, 0:t-1}),\sigma_{i, t}(\theta, h_{\cdot, 0:t-1}))
\label{eq:return_distribution}
\end{equation}

where $h_{\cdot, 0:t-1}$ represents historical information. Omitting the subscripts of $r^\text{data}$, $\mu$ and $\sigma$ for conciseness, our loss function is:
\begin{equation}
Loss(\theta) = \sum_{i, t} \left[\log{\sigma(\theta)} + \frac{(r^{\text{data}} - \mu(\theta))^2}{2 \sigma^2(\theta)} \right]
\label{eq:loss_function}
\end{equation}

\subsection{Model Architecture}
Our model can be divided into two parts: Adaptive Data Conditioning Layer and Spatial-Temporal Transformer.

\subsubsection{Adaptive Data Conditioning Layer}
To mitigate the over-reliance on certain high market value assets (e.g., Kweichow Moutai in the Chinese A-shares market), we draw on the basic concept of Dropout \cite{srivastava2014dropout} and propose a Market Value Dropout method. This method randomly drops training assets based on their market value to enhance model robustness and improve performance on the test set:
\begin{equation}
\begin{aligned}
&mask \sim \text{Multinomial}(\lfloor Np \rfloor, \text{Softmax}(\frac{\text{MarketValue}}{\tau})) \\
&MarketValueDropout(r^\text{data}, p) = \frac{1}{1 - p} \cdot mask \cdot r^\text{data}
\end{aligned}
\label{eq:market_value}
\end{equation}

where $p$ represents the probability of dropout and $\tau$ represents the temperature coefficient. The term $\frac{1}{1 - p}$ is used to scale the values proportionally, ensuring that the range of data remains consistent during training and inference.

Due to various reasons, financial data may be missing at certain time points, such as when trading is suspended or exchanges temporarily close. Commonly, methods like forward filling or interpolation are used to supplement missing data to aid model training. However, the absence of data itself can also be informative; for instance, a stock suspension might be due to significant corporate disclosures or serious company issues. Directly filling these gaps with fixed rules could result in a loss of this information.
XGBoost \cite{chen2016xgboost} handles missing values by splitting the missing samples into left and right branches, calculating their loss, and retaining the split direction that minimizes overall loss. Inspired by this idea, we allow the model to adaptively fill in missing values, as follows:
\begin{equation}
\begin{aligned}
ResMissingValue(x) = 
\begin{cases}
w_1, & \text{\textit{if} } x \text{\textit{ is missing}} \\
w_2 \cdot x + x, & \text{\textit{otherwise}}
\end{cases}
\end{aligned}
\label{eq:res_missing_value}
\end{equation}

 where $w_1$ and $w_2$ are trainable parameters. In cases where data is not missing, inspired by ResNet \cite{he2015deep}, we add a residual connection to the non-missing values to prevent network degradation since zero mapping is easier to train than identity mapping.

\subsubsection{Spatial-Temporal Transformer}

When forecasting stock returns, it is common to treat all stocks within an index as a single batch for gradient descent. However, this approach may overlook the influence between these stocks. For example, small-cap stocks may rise following a surge in large-cap stocks within the same industry, a phenomenon known as the "leader effect" within the sector.
Previous researchers have mostly conducted studies based on naturally divided industries \cite{helong2019survey}. Stocks grouped by natural industry divisions can indeed reflect most of the interactions between stocks. For instance, it is intuitive to believe that Apple, Microsoft, and Google would move in tandem. However, there are numerous, hard-to-identify interactions between stocks. To address this, we propose a low-rank attention-based asset encoder-decoder to capture the complex non-linear interactions between different assets, compressing high-dimensional assets into a low-dimensional space to enhance the prediction performance for asset families. Low-rank attention is used to capture the \textbf{"spatial"} features across assets, while the Transformer encoder-decoder captures the \textbf{"temporal"} features in asset time series. For data $E \in \mathbb{R}^{n \times d}$ with $d$ features and $n$ dimensions, we have
\begin{equation}
LowRankAttn(E, \theta) = \text{Softmax} \left( \frac{W^Q_\theta E^T}{\sqrt{d}} \right) E
\label{eq:low_rank_attention}
\end{equation}

where $W^Q_\theta \in \mathbb{R}^{k \times d}$ is a trainable weight matrix, and $k$ represents the dimension of the low-dimensional latent asset space. For example, we can set $k = 11$ according to the Global Industry Classification Standard (GICS), stocks can be divided into 11 sectors, suggesting that there are 11 unobservable latent assets that determine the direction of the entire market. However, different values of $k$ can also be explored to optimize performance.

Using a low-rank Attention Encoder, high-dimensional asset data is projected into a lower-dimensional latent space, where a Transformer processes the encoded historical information of the hidden assets to generate predictions. The low-rank Attention Decoder then maps the hidden assets back to the original asset dimensions. This dimensionality reduction and recovery process captures the complex relationships between different assets, reflecting the influence of market and industry dynamics on individual stocks. By predicting the distribution parameters $\mu(\theta)$ and $\sigma(\theta)$, and sampling from the resulting distribution, we obtain forecasts for asset returns. Thanks to the Transformer Decoder, the model leverages a masking mechanism for parallel training, while during inference, predictions are generated auto-regressively to create complete forecast trajectories.

\section{RL for hedging}
\subsection{Derivatives Hedging Theory}
Consider a financial market with discrete time $t \in \{0,1,2,\dots,T\}$, and define the probability space $(\Omega, \mathcal{F}, P)$, where $\mathcal{F}(t)$ is a $\sigma$-algebra satisfying $\mathcal{F}(1) \subset \mathcal{F}(2) \subset \dots \mathcal{F}(T) = \mathcal{F}$. Let $S_t \in \mathbb{R}^n$ be the positive-valued asset price process adapted to $\mathcal{F}(t)$. Assume that the derivative's payoff at time $T$ is $V_T$, which is $\mathcal{F}(T)$-measurable, and our goal is to hedge this derivative. Notice that we do not assume an equation for $S_t$ here, but we define
\begin{equation}
S_t = f(S_{t-1}, \epsilon_t)
\label{eq:state_transition}
\end{equation}

where $f$ is the transition equation for $S_t$, and $\epsilon_t$ is a random variable adapted to $\mathcal{F}(t)$. Using the prediction model from Chapter 2, we can write it as
\begin{equation}
S_t = f_\theta^{NN}(S_{t-1}, \epsilon_t)
\label{eq:nn_state_transition}
\end{equation}

where $\theta$ represents the parameters of the neural network.

At the initial time, we short the derivative to obtain funds $V_0$, and at the terminal time $T$, we face a derivative payment of $-V_T$. We need to use a trading strategy at $t \in \{0,1,2,\dots,T\}$ to hedge the portfolio value change caused by the terminal payment. This process is called hedging. Consider a strategy $\delta_t = \{\delta_{1,t}, \delta_{2,t}, \dots, \delta_{n,t}\} \in \mathbb{R}^n$ that is adapted to $\mathcal{F}(t)$ and represents the position in the underlying asset at time $t$. We define $\delta_{-1} = \delta_T = 0$, meaning we do not hold any positions before the start or after the end of the trading period. Assume all trades are self-financing (i.e., no external funds are injected into our portfolio, and its value changes solely due to the market and strategy). Thus, we define
\begin{equation}
(\delta \otimes s)_T = \sum_{t=0}^{T} \delta_t \cdot (S_{t+1} - S_t).
\label{eq:delta_summation}
\end{equation}

Then, at the terminal time, the total value of our portfolio $PV_T$ is
\begin{equation}
PV_T = -V_T + V_0 + (\delta \otimes s)_T.
\label{eq:pv_equation}
\end{equation}

In the real world, almost all trades are affected by transaction fees, commissions, slippage, and other cost factors, which often directly impact the investor's returns and the effectiveness of the strategy. Therefore, we consider trading friction costs. Let the switching cost from time $t-1$ to time $t$ be $Cost_t$, given by
\begin{equation}
Cost_t = h(\delta_{t-1}, \delta_t),
\label{eq:cost_function}
\end{equation}

where $h(\delta_{t-1}, \delta_t)$ represents the trading friction function for changing the position from $\delta_{t-1}$ to $\delta_t$. Generally, we assume that this cost is linear, i.e.,
\begin{equation}
Cost_t = c S_t \cdot |\delta_{t-1} - \delta_t|,
\label{eq:cost_t}
\end{equation}

where $c$ is the linear coefficient for the transaction cost rate. The total trading cost $C_T$ is then
\begin{equation}
C_T = \sum_{t=0}^{T} Cost_t = \sum_{t=0}^{T} c S_t \cdot |\delta_{t-1} - \delta_t|.
\label{eq:total_cost}
\end{equation}

Considering transaction costs, the total value of the portfolio at the terminal time $PV_T$ is
\begin{equation}
PV_T = -V_T + V_0 + (\delta \otimes s)_T - C_T 
= -V_T + V_0 + \sum_{t=0}^{T} \delta_t \cdot (S_{t+1} - S_t) 
- \sum_{t=0}^{T} c S_t \cdot |\delta_{t-1} - \delta_t|.
\label{eq:portfolio_value}
\end{equation}

The hedging problem can be formalized as a control problem, i.e.,
\begin{equation}
\max_{\substack{t = 0, 1, \dots, T-1 \\ \delta_t \in \Delta_t}} U(PV_T).
\label{eq:hedging_objective}
\end{equation}

where $\Delta_t$ represents the feasible set of positions for $\delta_t$, and $U(x)$ represents the objective function based on risk and return preferences. In the derivatives hedging task, we want the total value of the portfolio at the terminal time to be as close to 0 as possible. Therefore, following Goudenege and Molent\cite{goudenege2024backward}, we define the objective function as
\begin{equation}
U(x) = -x^2.
\label{eq:utility_function}
\end{equation}

which forces the agent to use the underlying asset to hedge the payments caused by the derivative's settlement at maturity as effectively as possible.

\subsection{Hedging in the Context of Reinforcement Learning}

We abstract the hedging problem into a reinforcement learning problem and provide the specific meanings of state, action, reward, and state transition. At time $t=0$, the agent enters the short position in the derivative and receives some funds, which originate from shorting the derivative. Taking a European call option as an example, these funds come from selling the option and receiving the option premium. We define the state observed by the agent at time $t$ as $s_t$, the underlying asset price as $S_t$, the derivative contract's maturity time as $T$, and the strike price of the derivative as $K$. Then, we have
\begin{equation}
s_t = \left( \frac{S_t}{K}, T - t, a_{t-1} \right).
\label{eq:state_representation}
\end{equation}

where $a_t$ represents the action taken by the agent at time $t$, corresponding to the position in the underlying asset $\delta_t$ as mentioned in Section 3.1.

The reward $r_t$ is defined as
\begin{equation}
r_t = 
\begin{cases}
0, & \text{if } 0 \leq t < T, \\
-PV_T^2, & \text{if } t = T.
\end{cases}
\label{eq:reward_function}
\end{equation}

where $PV_T$ is the total value of the portfolio at the terminal time, as shown in Equation \ref{eq:portfolio_value}. This definition causes the issue of sparse rewards. In reinforcement learning, reward signals are typically propagated recursively from the termination point backwards, as shown before. However, if we do not set rewards to be sparse, the cost of obtaining the portfolio at times $0 \leq t < T$ becomes quite high: either we train using real data, which leads to insufficient training data, or we generate the data for the underlying asset while also generating the derivative value data, which places too strict conditions on the generative model \cite{cohen2023arbitrage}, or we need to assume the motion equation of the underlying asset price, which results in poor backtest performance when the agent transitions to real data. Therefore, we introduce a temporal difference error and generalized advantage estimation based on $\lambda$-returns to improve the agent's performance in sparse rewards settings.

Let $T$ represent the state transition operator, such that $T(s_t, a) = s_{t+1}$. Then, we have
\begin{equation}
T\left( \left( \frac{S_t}{K}, T - t, a_{t-1} \right), a_t \right) = \left( \frac{f(S_t, \epsilon_{t+1})}{K}, T - t - 1, a_t \right).
\label{eq:state_transition}
\end{equation}

where $f$ is the transition function for $S_t$, as in Equation \ref{eq:nn_state_transition}, and $\epsilon_{t+1}$ is a random variable.

\subsection{Behavior Cloning - Proximal Policy Optimization}

\begin{figure}[ht]
    \centering
    \includegraphics[width=0.8\textwidth]{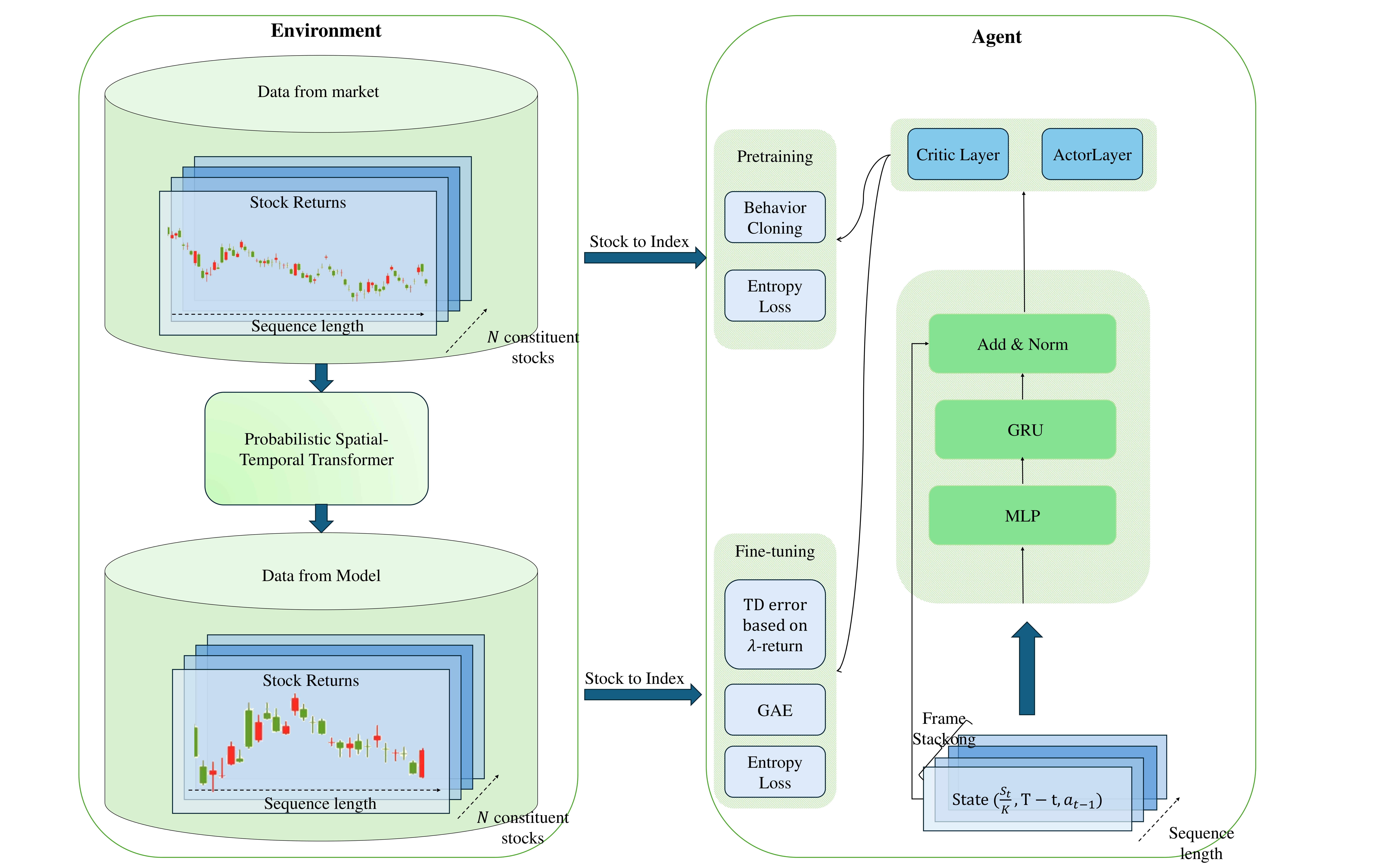}
    \caption{Behavior Cloning - Proximal Policy Optimization}
    \label{fig:rl_arch}
\end{figure}

We construct the training environment using a spatiotemporal attention-based probabilistic financial time series prediction Transformer and real market data. The training employs a pretraining-finetuning architecture. During the finetuning phase, we improve the Proximal Policy Optimization (PPO) algorithm by incorporating $n$-step temporal difference errors to enhance performance in sparse reward scenarios. Additionally, we introduce frame stacking and gated recurrent units (GRUs) to improve the agent's performance in hedging trades, as shown in Figure \ref{fig:rl_arch}.

\subsubsection{Pretraining Based on Behavior Cloning}

To improve data efficiency and guide the agent in effective exploration, we use expert data and behavior cloning pretraining. This integrates the expert's strategy directly into the agent's learning process, boosting the agent's capabilities at the beginning of training. The expert data is derived from real historical market data. For a European call option, we describe the process of constructing the expert dataset. For a geometric Brownian motion-driven underlying asset price model, assuming a constant risk-free rate $r_f$, we have

\begin{equation}
\frac{dS(t)}{S(t)} = \mu dt + \sigma dW(t),
\label{eq:gbm}
\end{equation}

where $\mu$ and $\sigma$ are constants. Theoretically, the price of a European call option expiring at time $T$ with strike price $K$ is

\begin{equation}
c = S(0) N(d_1) - K e^{-r_f T} N(d_2),
\label{eq:call_price}
\end{equation}

\begin{equation}
d_1 = \frac{\ln\left( \frac{S(0)}{K} \right) + \left( r_f + \frac{\sigma^2}{2} \right) T}{\sigma \sqrt{T}}, \quad 
d_2 = d_1 - \sigma \sqrt{T},
\label{eq:d1_d2}
\end{equation}

where $N(\cdot)$ represents the cumulative distribution function of the standard normal distribution. In practice, the observed market price of the option is $c_{market}$, and in Equation \ref{eq:d1_d2}, only the volatility $\sigma$ is unknown. We can invert this to solve for the implied volatility $\sigma_{market}$ from $c_{market}$, which is referred to as the implied volatility. Financially, this value represents the market's expectation of future volatility.

The market's implied Delta value $\Delta_{market}$ is given by

\begin{equation}
\Delta_{market} = \frac{\partial c}{\partial S(0)} = N\left( \frac{\ln\left( \frac{S(0)}{K} \right) + \left( r_f + \frac{\sigma_{market}^2}{2} \right) T}{\sigma_{market} \sqrt{T}} \right).
\label{eq:delta_market}
\end{equation}

In the case where the underlying asset price follows Equation \ref{eq:gbm}, and allowing for continuous trading with no transaction fees, holding a position corresponding to the Delta value perfectly hedges the short position in the European call option. This is because Delta is the derivative of the option price with respect to the underlying asset price. The $\Delta_{market}$ calculated using the implied volatility reflects the market's sensitivity to changes in the underlying asset price, and we use $\Delta_{market}$ as the expert action. During pretraining, the agent learns the expert's actions and strategies. In the experimental phase, we also use it as a benchmark strategy for comparison.

In the pretraining phase, we use the maximum likelihood loss to assist the agent in learning the expert strategy. For a state $s$, we represent the agent's action distribution as $a \sim N(\mu_\theta^{NN}(s), \sigma_\theta^{NN2}(s))$, where $\mu_\theta^{NN}$ and $\sigma_\theta^{NN}$ are parameterized mean and variance functions formed by a neural network. We denote the expert dataset containing $N_{expert}$ state-action pairs as $\{s_i^{expert}, a_i^{expert}\}_{1 \leq i \leq N_{expert}}$. During batch learning, we propose the behavior cloning loss based on maximum likelihood:

\begin{equation}
\text{Loss}_{expert}(\theta) = \frac{1}{N_{batch}} \sum_{i=1}^{N_{batch}} \left[ \frac{1}{2} \ln(2\pi \sigma_\theta^{NN2}(s_i^{expert})) + \frac{1}{2 \sigma_\theta^{NN2}(s_i^{expert})} \left( \mu_\theta^{NN}(s_i^{expert}) - a_i^{expert} \right)^2 \right].
\label{eq:loss_expert}
\end{equation}

where $N_{batch}$ is the batch size randomly sampled from the expert dataset $\{s_i^{expert}, a_i^{expert}\}_{1 \leq i \leq N_{expert}}$. To prevent the model from falling into local minima during pretraining and to ensure the agent's exploratory behavior during finetuning, we propose an entropy loss $\text{Loss}_{entropy}(\theta)$:

\begin{equation}
\text{Loss}_{entropy}(\theta) = -\frac{1}{2 N_{batch}} \sum_{i=1}^{N_{batch}} \ln(2\pi e \sigma_\theta^{NN2}(s_i^{expert})).
\label{eq:loss_entropy}
\end{equation}

The parameter update during the pretraining phase is represented as

\begin{equation}
\theta \leftarrow \theta - \eta \nabla \left( \text{Loss}_{expert}(\theta) + \lambda_{entropy} \text{Loss}_{entropy}(\theta) \right).
\label{eq:theta_update}
\end{equation}

where $\eta$ represents the learning rate and $\lambda_{entropy} > 0$ represents the trade-off between fitting the expert action and maintaining exploratory behavior in the agent's strategy.

The pretraining algorithm based on behavior cloning is shown in Algorithm 1.

\begin{algorithm}[ht]
\caption{Behavior Cloning Pretraining for the Agent}
\KwIn{Randomly initialized parameters $\mu_\theta^{NN}$ and $\sigma_\theta^{NN}$ for the agent's strategy, expert dataset $\{s_i^{expert}, a_i^{expert}\}_{1 \leq i \leq N_{expert}}$}
\KwOut{Agent with integrated expert strategy}
\For{$i \in \{1, 2, \dots, n_{epoch}\}$} {
    Sample a batch of $N_{batch}$ expert state-action pairs from $\{s_i^{expert}, a_i^{expert}\}_{1 \leq i \leq N_{expert}}$ \;
    Compute the maximum likelihood loss:
    \[
    \text{Loss}_{expert}(\theta) = \frac{1}{N_{batch}} \sum_{i=1}^{N_{batch}} \left[ \frac{1}{2} \ln(2\pi \sigma_\theta^{NN2}(s_i^{expert})) + \frac{1}{2 \sigma_\theta^{NN2}(s_i^{expert})} \left( \mu_\theta^{NN}(s_i^{expert}) - a_i^{expert} \right)^2 \right].
    \] \\
    Compute the entropy loss:
    \[
    \text{Loss}_{entropy}(\theta) = -\frac{1}{2 N_{batch}} \sum_{i=1}^{N_{batch}} \ln(2\pi e \sigma_\theta^{NN2}(s_i^{expert})).
    \] \\
    Compute the pretraining loss:
    \[
    \text{Loss}_{pretrain} = \text{Loss}_{expert}(\theta) + \lambda_{entropy} \text{Loss}_{entropy}(\theta).
    \] \\
    Update parameters: $\theta \leftarrow \theta - \eta \nabla \text{Loss}_{pretrain}(\theta)$.
}
\KwOut{Trained agent with integrated expert strategy}
\end{algorithm}

\subsubsection{Fine-Tuning Based on Proximal Policy Optimization with Recurrent Networks}

In the fine-tuning phase, we apply reinforcement learning algorithms to fine-tune the agent's policy distribution. Although the agent's policy distribution aligns with the expert policy during the pretraining phase, the trading strategy based on the market-implied Delta values is not necessarily the optimal strategy in the hedging environment. Firstly, the market may face unforeseen black swan events, such as natural disasters or macroeconomic policy changes, which are unpredictable. Additionally, there are various transaction costs in the market, including stamp duties, commissions, and slippage, which are not considered in theoretical models. By applying reinforcement learning algorithms, we can improve the agent's performance out-of-sample by leveraging the exploration advantages of these algorithms. We have improved the Proximal Policy Optimization (PPO) algorithm to make the agent's training more efficient for the problem we are studying.

Although we discussed value-function-based methods and policy gradient methods separately in Section 3.2, in modern deep reinforcement learning frameworks, we typically use the Actor-Critic (AC) framework, which estimates the value function while optimizing the policy distribution. The role of the actor is to select actions based on the current state. It outputs a probability distribution over actions based on the current policy. The actor's goal is to maximize the cumulative reward, and it updates its policy based on feedback from the critic. The critic's task is to accurately assess the value of the agent's action in a given state. Below, we detail the update algorithms for the actor and critic.

For the actor's learning, we primarily use the policy gradient method. Generally, when using the policy gradient method to optimize the agent's policy, we can replace \(G_0 (\tau)\) with \(\Phi(\tau)\) to assign reasonable scores to the policy. In our hedging trading case, if the rewards are sparse, the real-time reward will be zero for any action taken at intermediate states, which is not conducive to optimizing the agent’s policy. Therefore, during training, we tend to assign a suitable score to each action the agent might take. We expand the function \(\Phi(\tau)\) that scores the entire trajectory, writing it as a function of state and action, denoted \(\Phi(s,a)\). Schulman et al. \cite{schulman2015high} propose several definitions for \(\Phi(s,a)\). Noting the definitions of the state value function \(V_{\pi}(s)\) and the action-value function \(Q_{\pi}(s,a)\), we can define \(\Phi(s,a) = Q_{\pi}(s,a)\), using the action-value function to evaluate the value of state-action pairs. We can also define \(\Phi(s,a) = Q_{\pi}(s,a) - V_{\pi}(s)\), which is known as the advantage function. Subtracting the \(V_{\pi}(s)\), which is independent of \(a\), from \(Q_{\pi}(s,a)\) is reasonable as it not only aligns with the motivation of assigning reasonable scores to each action but also reduces variance. Furthermore, we can define \(\Phi(s,a) = r_t + \gamma V_{\pi}(s') - V_{\pi}(s)\), where \(s'\) represents the next state, and in this case, \(\Phi(s,a)\) is the temporal difference error. When \(V_{\pi}\) and \(Q_{\pi}\) are known, the policy gradient guarantees convergence to the optimal policy \cite{schulman2015high}.

However, we do not know the true \(V_{\pi}(s)\), and estimates of the value function always have bias. We denote the parameterized estimate of the value function as \(V_{\pi}^{\theta}\). One estimate for the advantage function is:

\begin{equation}
\bar{\Phi}(s,a) = r_t + \gamma V_{\pi}^{\theta}(s') - V_{\pi}^{\theta}(s).
\label{eq:advantage_function}
\end{equation}

However, this introduces bias because the true \(V_{\pi}\) is replaced by the estimate \(V_{\pi}^{\theta}\). Another estimate using Monte Carlo methods is:

\begin{equation}
\bar{\Phi}(s,a) = r_t + \gamma r_{t+1} + \cdots.
\label{eq:advantage_function_expanded}
\end{equation}

Since this method requires traversing the entire trajectory, it introduces variance. We use the idea of \(\lambda\)-returns to balance bias and variance and define the \(k\)-step temporal difference error. By selecting \(k\), we can strike a balance between the Monte Carlo estimate and the temporal difference:

\begin{equation}
TD_t^{(k)} = -V_{\pi}^{\theta}(s_t) + r_t + \gamma r_{t+1} + \cdots + \gamma^{k-1} r_{t+k-1} + \gamma^k V_{\pi}^{\theta}(s_{t+k}).
\label{eq:td_k_step}
\end{equation}

Using all \(k\)-step temporal difference errors with exponential weighting, we get the generalized advantage estimate \(\bar{\Phi}(s,a)\). , we have:

\begin{equation}
\begin{aligned}
\bar{\Phi}(s,a) &= (1-\lambda) \left( \sum_{i=1}^{\infty} \lambda^{i-1} TD_t^{(i)} \right) \\
&= (1-\lambda) (TD_t^{(1)} + \lambda TD_t^{(2)} + \lambda^2 TD_t^{(3)} + \cdots) \\
&= (1-\lambda) (TD_t^{(1)} (1 + \gamma + \gamma^2 + \cdots) + \gamma TD_{t+1}^{(1)} (\gamma + \gamma^2 + \cdots) + \cdots) \\
&= \sum_{i=0}^{\infty} (\lambda \gamma)^i TD_{t+i}^{(1)}.
\end{aligned}
\label{eq:gae_final}
\end{equation}

Since our hedging trading environment is a finite-horizon decision problem, all rewards for \(t > T\) are set to zero, and thus the final step of Equation \ref{eq:gae_final} is truncated. Then, we can improve the agent's performance in the hedging task by gradient ascent, written in terms of the gradient \(g\):

\begin{equation}
g = E_t \left( \nabla \ln \pi_{\theta}(a_t | s_t) \bar{\Phi}(s_t, a_t) \right).
\label{eq:policy_gradient}
\end{equation}

The negative sign is to reverse the direction of the gradient. Note that since our sample trajectories are collected under the policy \(\pi_{\theta}(a|s)\), once \(\theta\) is updated, the policy \(\pi_{\theta}(a|s)\) changes, and previously collected sample trajectories can no longer be used for the current agent’s update, leading to lower data efficiency in policy gradient methods. To address this, Schulman et al. \cite{schulman2015high} proposed an approximate algorithm using importance sampling. The gradient is multiplied by the ratio of the current policy to the historical policy, allowing data collected by other policies to be used for the current policy's update. The objective function is:

\begin{equation}
\max_{\theta} E_t \left[ \frac{\pi_{\theta}(a_t | s_t)}{\pi_{\theta_{\text{old}}}(a_t | s_t)} \bar{\Phi}(s_t, a_t) - \beta \text{KL} \left[ \pi_{\theta_{\text{old}}}(\cdot | s_t), \pi_{\theta}(\cdot | s_t) \right] \right],
\label{eq:ppo_objective}
\end{equation}

where KL denotes the Kullback-Leibler divergence, \(\pi_{\theta_{\text{old}}}\) represents the historical policy, and \(\beta\) controls the difference between the current and historical policy distributions. Empirically, a truncated optimization strategy works better. The loss function used to update the actor is:

\begin{equation}
L^{\text{actor}}(\theta) = -E_t \left[ \frac{\pi_{\theta}(a_t | s_t)}{\pi_{\theta_{\text{old}}}(a_t | s_t)} \bar{\Phi}(s_t, a_t) \right].
\label{eq:actor_loss}
\end{equation}

The critic's learning is primarily based on value function estimation, which can be written in the form of a loss function:

\begin{equation}
\text{Loss}^{\text{VF}}(\theta) = \left( V^{\theta}(s_t) - V_t^{\text{target}} \right)^2
\label{eq:value_function_loss}
\end{equation}

In PPO, Schulman et al. \cite{schulman2017proximal} use temporal difference error for training, where $V_t^{\text{target}} = r_t + \gamma V^{\theta}(s_{t+1})$. Similarly, this method introduces bias. Additionally, in our sparse reward environment, using temporal difference error to train $V^{\theta}$ is difficult to converge, because the reward signal only appears after the entire transaction is complete. In this case, for $0 \leq t < T$, $r_t = 0$, and temporal difference error cannot provide effective reward signals. Only when $r_T \neq 0$ does $V^{\theta}$'s parameter receive effective updates. The Monte Carlo algorithm, where $V_t^{\text{target}} = r_t + \gamma r_{t+1} + \cdots + \gamma^{(T-t)} r_T$, although unbiased for $V_t^{\text{target}}$ when $0 \leq t < T$, has a high variance. To strike a balance between variance and bias, we use the idea of $\lambda$-returns to estimate $V_t^{\text{target}}$, as follows:

\begin{equation}
\begin{aligned}
V_t^{\text{target}} &= r_{t+1} + \gamma r_{t+2} + \cdots \\
&= r_{t+1} + \gamma V_{\pi}^{\theta}(s_{t+1}) - V_{\pi}^{\theta}(s_t) - \gamma V_{\pi}^{\theta}(s_{t+1}) + V_{\pi}^{\theta}(s_t) + \gamma r_{t+2} + \cdots \\
&= V_{\pi}^{\theta}(s_t) + \text{TD}_t^{(1)} - \gamma V_{\pi}^{\theta}(s_{t+1}) + \gamma r_{t+2} + \cdots \\
&= V_{\pi}^{\theta}(s_t) + \text{TD}_t^{(1)} + \gamma \left( -V_{\pi}^{\theta}(s_{t+1}) + \gamma V_{\pi}^{\theta}(s_{t+2}) + r_{t+2} - \gamma V_{\pi}^{\theta}(s_{t+2}) \right) + \cdots \\
&= V_{\pi}^{\theta}(s_t) + \text{TD}_t^{(1)} + \gamma \text{TD}_{t+1}^{(1)} + \cdots \\
&= V_{\pi}^{\theta}(s_t) + \sum_{n=0}^{k-1} \gamma^n \text{TD}_{t+n}^{(1)} \\
&= V_{\pi}^{\theta}(s_t) + \bar{\Phi}(s_t, a_t)
\end{aligned}
\label{Vt_target}
\end{equation}

Using this method, we can more accurately estimate $V_t^{\text{target}}$ to assist in updating $V^{\theta}$ and avoid the convergence issues caused by sparse rewards. Furthermore, this method is computationally efficient, as we can quickly compute $V_t^{\text{target}}$ after calculating the generalized advantage estimate $\bar{\Phi}(s_t, a_t)$, thus improving training efficiency.

In the theoretical model, the underlying asset price is assumed to follow a Markov process, which deviates from real-world financial markets. Real-world financial markets often have complex dynamic characteristics, with many factors influencing market price fluctuations, and these factors generally have long-term dependencies. Therefore, relying solely on the current state to make decisions may fail to capture the true dynamic characteristics of the market, limiting the model's prediction accuracy and the strategy's effectiveness.

Following the approach of Massahi and Mahootchi \cite{massahi2024deep}, we use recurrent neural networks (RNNs) to capture temporal dependencies in the underlying asset features. Specifically, we employ Gated Recurrent Units (GRU) \cite{chung2014empirical} to capture temporal features while avoiding issues like vanishing or exploding gradients. Additionally, we use frame stacking \cite{liu2025combining}, stacking historical states as input to the agent. This enhances the agent's "memory" during decision-making, enabling it to gain a more comprehensive understanding of the changes and fluctuations in the underlying asset.

At the same time, considering the importance of the current state, to prevent deep network degradation \cite{he2015deep}, we also apply a residual connection for the most recent state to the agent's output. This ensures that the model remains stable during fine-tuning and maintains information flow during training, ensuring that important current state information is effectively passed on. With the improvements of frame stacking, GRU, and residual connections, the agent can effectively capture both long-term and short-term temporal dependencies when taking actions, avoiding information loss.

The fine-tuning algorithm with recurrent PPO is as follows:

\begin{algorithm}[htbp]
\caption{Fine-Tuning with Recurrent PPO for Hedging}
\label{alg:ppo_recurrent_finetune}
\KwIn{Pre-trained agent policy $\pi_{\theta_1}$, value function parameterized by $V^{\theta_2}$, derivative hedging environment $env$ constructed with spatio-temporal attention-based probabilistic financial time series prediction Transformer}
\KwOut{Hedging agent trained for $n_{\text{epoch}}$ epochs}

\For{$i \in \{1, 2, \dots, n_{\text{epoch}}\}$}{
    Explore environment with policy $\pi_{\theta_1}$ to obtain $n_{\text{path}}$ trajectories $\{s_i, a_i, r_i\}_{1 \leq i \leq n_{\text{path}}}$\;
    $\theta_{\text{old}} \gets \theta_1$\;
    \For{$j \in \{1, 2, \dots, n_{\text{actor\_update}}\}$}{
        Compute Generalized Advantage Estimation (GAE): 
        \[
        \bar{\Phi}_{\theta_2}(s_i, a_i) \gets \sum_{i=0}^{T} (\lambda \gamma)^i \text{TD}_{t+i}^{(1)} 
        \] \;
        Compute clipped policy gradient loss:
        \[
        \text{Loss}^{\text{CLIP}} (\theta_1) \gets \frac{1}{n_{\text{path}}} \sum_{t=1}^{n_{\text{path}}} \min \left( \frac{\pi_{\theta_1}(a_t|s_t)}{\pi_{\theta_{\text{old}}}(a_t|s_t)} \bar{\Phi}_{\theta_2}(s_t, a_t), \text{clip} \left( \frac{\pi_{\theta_1}(a_t|s_t)}{\pi_{\theta_{\text{old}}}(a_t|s_t)}, 1-\epsilon, 1+\epsilon \right) \bar{\Phi}_{\theta_2}(s_t, a_t) \right)
        \] \;
        Compute entropy loss using Equation \ref{eq:loss_entropy} \;;
        Update $\theta_1 \gets \theta_1 - \eta \nabla \left( \text{Loss}^{\text{CLIP}}(\theta_1) + \lambda_{\text{entropy}} \text{Loss}_{\text{entropy}} (\theta_1) \right)$ \;
    }
    \For{$j \in \{1, 2, \dots, n_{\text{critic\_update}}\}$}{
        Compute value function loss:
        \[
        \text{Loss}^{\text{VF}}(\theta_2) \gets \sum_{t=1}^{n_{\text{path}}} \bar{\Phi}_{\theta_2}^2 (s_t, a_t)
        \] \
        Update $\theta_2 \gets \theta_2 - \eta \nabla \text{Loss}^{\text{VF}} (\theta_2)$ \;
    }
}
\textbf{Output:} $\pi_{\theta_1}$
\end{algorithm}

\section{Numerical Experiments Based on the CSI 300 Index and the S\&P 500 Index}

In the study of derivative hedging and asset pricing, stock index futures and stock index options are important financial derivatives that have received widespread attention. The CSI 300 Index (China Securities Index 300, CSI 300) and the S\&P 500 Index (Standard \& Poor's 500, S\&P 500) are two prominent market indices, which have significant influence and representativeness in the Chinese and U.S. financial markets, respectively. Therefore, investigating the market behaviors of these two indices and the effectiveness of derivative hedging strategies is not only essential for understanding price fluctuation patterns in different market structures, but also provides more scientifically grounded hedging methods and strategies for market participants.

This chapter aims to explore the prediction of constituent stocks and derivative hedging issues for the CSI 300 Index and the S\&P 500 Index through numerical experiments. Firstly, considering the differences and similarities between these two markets, the chapter will apply a spatiotemporal attention-based probabilistic financial time-series prediction Transformer model. This model will be used to fit the price movements of both indices based on historical market data and generate future price prediction paths. The experiments will demonstrate the model’s ability to effectively capture the complex nonlinear relationships between markets and assess the quality of the predicted data, thus providing a foundation for optimizing derivative hedging strategies. Additionally, the accuracy of the derivative pricing method proposed in Section 2.4 will be evaluated. 

Next, the chapter will collect corresponding index option data and, combining reinforcement learning methods, address the hedging problem of index options for the CSI 300 and S\&P 500 indices. The reinforcement learning agent will be trained and optimized, and the algorithm's performance will be evaluated using real market data. Through numerical experiments, the chapter will explore the performance of the behavior cloning-proximal policy optimization (BC-PPO) algorithm in different market environments and assess its effectiveness and robustness in real-world trading.

\subsection{Data Description}

This study collected all data for the CSI 300 Index and the S\&P 500 Index from January 4, 2010, to July 1, 2024, at a daily frequency. The data is divided into training, validation, and test sets based on time. The training set covers data from January 1, 2010, to September 29, 2023; the validation set includes data from September 30, 2023, to December 31, 2023; and the test set includes data from January 1, 2024, to June 30, 2024. The CSI 300 Index and constituent stock data were sourced from BaoStock (\url{www.baostock.com}), while the S\&P 500 Index and constituent stock data were obtained from Yahoo Finance (\url{finance.yahoo.com}).

Regarding derivatives, this study collected daily data for CSI 300 Index options and S\&P 500 Index options (S\&P 500 Index Options, SPX Options) during the corresponding period. Data for CSI 300 Index options came from the China Financial Futures Exchange (\url{www.cffex.com.cn}), while S\&P 500 Index options data was sourced from the Chicago Board Options Exchange (\url{www.cboe.com}). Since the CSI 300 Index options were launched relatively late, on December 23, 2019, this study only collected data from June 1, 2020, onward to avoid the abnormal market fluctuations caused by the launch of new options contracts.

Additionally, some deep in-the-money, deep out-of-the-money, and long-dated options had low trading volumes. This study excluded options contracts with zero trading volume, retaining only near-term call option contracts with 0.95 $\leq$ $S_0/K$ $\leq$ 1.05 and expirations within one month.

All experiments were conducted on WSL 2 (Ubuntu 22.04), with hardware configuration consisting of an Intel Core i5-8300H CPU and an NVIDIA GeForce GTX 1050 Ti (4GB VRAM) GPU. The deep learning framework used was PyTorch version 2.1.2 with CUDA 12.3.

\subsection{Hedging Task}

In the hedging task involving European call options on the CSI 300 and S\&P 500 indices, for all eligible options, the agent initially shorts the option at the market price at the start of the trading period. Subsequently, the agent is only allowed to trade the corresponding underlying asset (the market index) in long positions until the option expires, and all financial assets are settled at expiration. Under the assumption of rational counterparties, the payoff $V_T$ at expiration for the short position in the European call option is given by:

\begin{equation}
V_T = -\max(S_T - K, 0),
\end{equation}

\noindent where $S_T$ is the stock price on the expiration date, and $K$ is the strike price. In previous studies, such as that of Gao et al. \cite{gao2023deeper}, the state space only involves the underlying asset's price, and a single training session can only solve the hedging problem for options with a single strike price. In contrast, this study defines the state space in Equation \ref{eq:state_representation}, providing the strike price information to the agent, enabling it to handle options with multiple strike prices.

\subsection{Performance Metrics}

The performance evaluation of the hedging strategy is primarily based on the statistical metrics of the total portfolio value $PV_T$ at the terminal time, as shown in Equation \ref{eq:portfolio_value}. These include the average reward (avg\_r) representing the hedging error, the mean (avg\_PV) and standard deviation (std\_PV) of the terminal portfolio value. Let the terminal portfolio value for the $i$-th option under strategy $\pi$ be denoted as $PV_{T,i}(\pi)$, where $1 \leq i \leq n$. The metrics are defined as follows:

\begin{equation}
\label{eq:hedging_metrics}
\begin{aligned}
\text{avg\_r} = \frac{1}{n} \sum_{i=1}^{n} U(PV_{T,i}(\pi)) = \frac{1}{n} \sum_{i=1}^{n} -PV_{T,i}^2 (\pi), \\
\text{avg\_PV} = \frac{1}{n} \sum_{i=1}^{n} PV_{T,i}(\pi), \\
\text{std\_PV} = \sqrt{\frac{1}{n-1} \sum_{i=1}^{n} (PV_{T,i}(\pi) - \text{avg\_PV})^2}.
\end{aligned}
\end{equation}

\subsection{Baseline Algorithms}

To assess the hedging performance of the behavior cloning-proximal policy optimization (BC-PPO) algorithm, we compare it with the implied Delta computed from Equation \ref{eq:delta_market} and several baseline reinforcement learning algorithms, including Advantage Actor-Critic (A2C) \cite{huang2022a2c}, Proximal Policy Optimization (PPO) \cite{schulman2017proximal}, Deep Deterministic Policy Gradient (DDPG) \cite{lillicrap2015continuous}, Twin Delayed Deep Deterministic Policy Gradient (TD3) \cite{fujimoto2018addressing}, and Soft Actor-Critic (SAC) \cite{haarnoja2018soft}.

A2C uses only policy gradient methods for training, with the loss function defined as:

\begin{equation}
\label{eq:loss_a2c}
\text{Loss}^{A2C}(\theta) = \mathbb{E}_t \left[ \log \pi_\theta(a_t | s_t) \Phi_t \right],
\end{equation}

\noindent where $\Phi_t$ represents the advantage function. Unlike PPO, A2C can only use a single sample from the exploration phase for one gradient update and lacks entropy loss, which results in lower data efficiency.

As described in Section 3.3, PPO uses importance sampling to perform multiple updates to the policy based on exploration samples. The loss function for PPO is:

\begin{equation}
\label{eq:loss_ppo}
\text{Loss}^{PPO}(\theta) = -\mathbb{E}_t \left[ \min \left( \frac{\pi_\theta(a_t | s_t)}{\pi_{\theta_{\text{old}}}(a_t | s_t)} \Phi_t, \text{clip} \left( \frac{\pi_\theta(a_t | s_t)}{\pi_{\theta_{\text{old}}}(a_t | s_t)}, 1-\epsilon, 1+\epsilon \right) \Phi_t \right) \right],
\end{equation}

\noindent where the clipping mechanism prevents large updates to the policy and ensures stability.

DDPG differs from PPO and A2C by being a continuous action space reinforcement learning algorithm that uses a deterministic policy $\mu$ to collect samples and estimates the Q-network using one-step temporal difference errors. The loss function for DDPG is:

\begin{equation}
\label{eq:loss_q}
\text{Loss}^Q(\theta) = \mathbb{E}_t \left[ \left( r_t + \gamma Q'(s_{t+1}, \mu'(s_{t+1} | \theta^{\mu'})) - Q(s_t, a_t | \theta^Q) \right)^2 \right],
\end{equation}

\noindent where $Q$ and $Q'$ are the critic and target Q-networks, respectively, and $\mu'$ is the target policy. The policy network is updated using:

\begin{equation}
\label{eq:policy_gradient}
\nabla_{\theta^\mu} J = \mathbb{E}_t \left[ \nabla_a Q(s_t, \mu(s_t) | \theta^Q) \nabla_{\theta^\mu} \mu(s_t | \theta^\mu) \right].
\end{equation}

DDPG uses target networks to stabilize the training process. Based on DDPG, TD3 addresses the overestimation of Q-values in DDPG by using double Q-networks and incorporating delayed updates for both the policy and target Q-networks, with noise smoothing to improve stability in practical applications.

SAC, on the other hand, optimizes for both reward maximization and entropy maximization to promote exploration. Its loss functions are as follows:

\begin{equation}
\begin{aligned}
\text{Loss}^V(\psi) = \mathbb{E}_t \left[ \frac{1}{2} \left( V_\psi(s_t) - \mathbb{E}_{a_t} \left[ Q_\theta(s_t, a_t) - \log \pi_\phi(a_t | s_t) \right] \right)^2 \right], \\
\text{Loss}^Q(\theta) = \mathbb{E}_t \left[ \frac{1}{2} \left( Q_\theta(s_t, a_t) - r_t - \gamma \mathbb{E}_t \left[ V_\psi(s_{t+1}) \right] \right)^2 \right], \\
\text{Loss}^\pi(\phi) = \mathbb{E}_t \left[ \text{KL}\left( \pi_\phi(\cdot | s_t) \| \frac{\exp(Q_\theta(s_t, \cdot))}{Z_\theta(s_t)} \right) \right].
\end{aligned}
\label{eq:loss_functions}
\end{equation}

\noindent where $\psi$ is the parameter for the value function, $\theta$ is the parameter for the action value function, and $\phi$ is the parameter for the policy distribution. $Z_\theta(s_t)$ is the normalization factor for the distribution.

Additionally, the implied Delta computed from Equation \ref{eq:delta_market} is used as a baseline strategy for comparison.

\subsection{Comparison with Baseline Algorithms Under Zero Transaction Costs}

This study compares the hedging performance of the Behavior Cloning-Proximal Policy Optimization (BC-PPO) algorithm with other reinforcement learning algorithms on options for the CSI 300 index and the S\&P 500 index. Additionally, as a reference, the hedging performance using the implied Delta calculated from the option prices is provided. The primary focus is on the hedging error, denoted by avg\_r, while the mean and standard deviation of the portfolio's terminal value also reflect the hedging strategy's preference for risk and return.

\begin{table}[h]
\centering
\caption{Hedging Error Comparison for CSI 300 Index Options under Zero Transaction Costs}
\label{tab:csi300_zero_tc}
\begin{tabular}{lccc}
\toprule
Model & avg\_r ($\times 10^{-3}$) & avg\_PV & std\_PV \\
\midrule
\textbf{Ours}           & \textbf{-0.5288}  & \textbf{4.2654}  & \textbf{22.5967} \\
Implied Delta  & -0.6017  & 3.6366  & 24.2591 \\
A2C            & -0.8053  & 2.7432  & 28.2441 \\
PPO            & -0.8469  & 2.1563  & 29.0223 \\
DDPG           & -12.5867 & -6.1977 & 112.0193 \\
TD3            & -17.6943 & -15.8930 & 132.0672 \\
SAC            & -6.1815  & -1.5487  & 78.6075 \\
\bottomrule
\end{tabular}
\end{table}

\begin{table}[h]
\centering
\caption{Hedging Error Comparison for S\&P 500 Index Options under Zero Transaction Costs}
\label{tab:sp500_zero_tc}
\begin{tabular}{lccc}
\toprule
Model & avg\_r ($\times 10^{-3}$) & avg\_PV & std\_PV \\
\midrule
\textbf{Ours}           & \textbf{-0.1748}  & \textbf{1.2726}  & \textbf{13.1599} \\
Implied Delta  & -0.1837  & 2.1457  & 13.3823 \\
A2C            & -2.6198  & -0.7277 & 51.1787 \\
PPO            & -1.0049  & 4.3841  & 31.3959 \\
DDPG           & -7.2950  & 25.4495 & 81.5313 \\
TD3            & -8.1440  & 19.6099 & 88.0876 \\
SAC            & -3.2490  & 5.2639  & 56.7565 \\
\bottomrule
\end{tabular}
\end{table}

Tables \ref{tab:csi300_zero_tc} and \ref{tab:sp500_zero_tc} present the hedging errors for different agents and implied Delta on CSI 300 index options and S\&P 500 index options. In the experiment with CSI 300 index options, the proposed algorithm achieved the smallest average hedging error and provided a higher portfolio return, indicating that there is an arbitrage opportunity in the market. The proposed algorithm successfully identified this arbitrage by hedging the option payoff. On the S\&P 500 index options, the proposed algorithm again achieved the smallest average hedging error, though the portfolio return differed somewhat from the implied Delta. 

DDPG, TD3, and SAC faced challenges in converging in sparse reward environments. Although DDPG achieved a high return in the S\&P 500 index option hedging, its hedging error was quite large, as the model failed to recognize the reward signals provided by the options.

To provide a comprehensive comparison of the algorithm's hedging ability, the distribution of the total portfolio value $PV_T$ at the terminal time and its Kernel Density Estimation (KDE) are plotted.

\begin{figure}[h]
\centering
\includegraphics[width=0.8\textwidth]{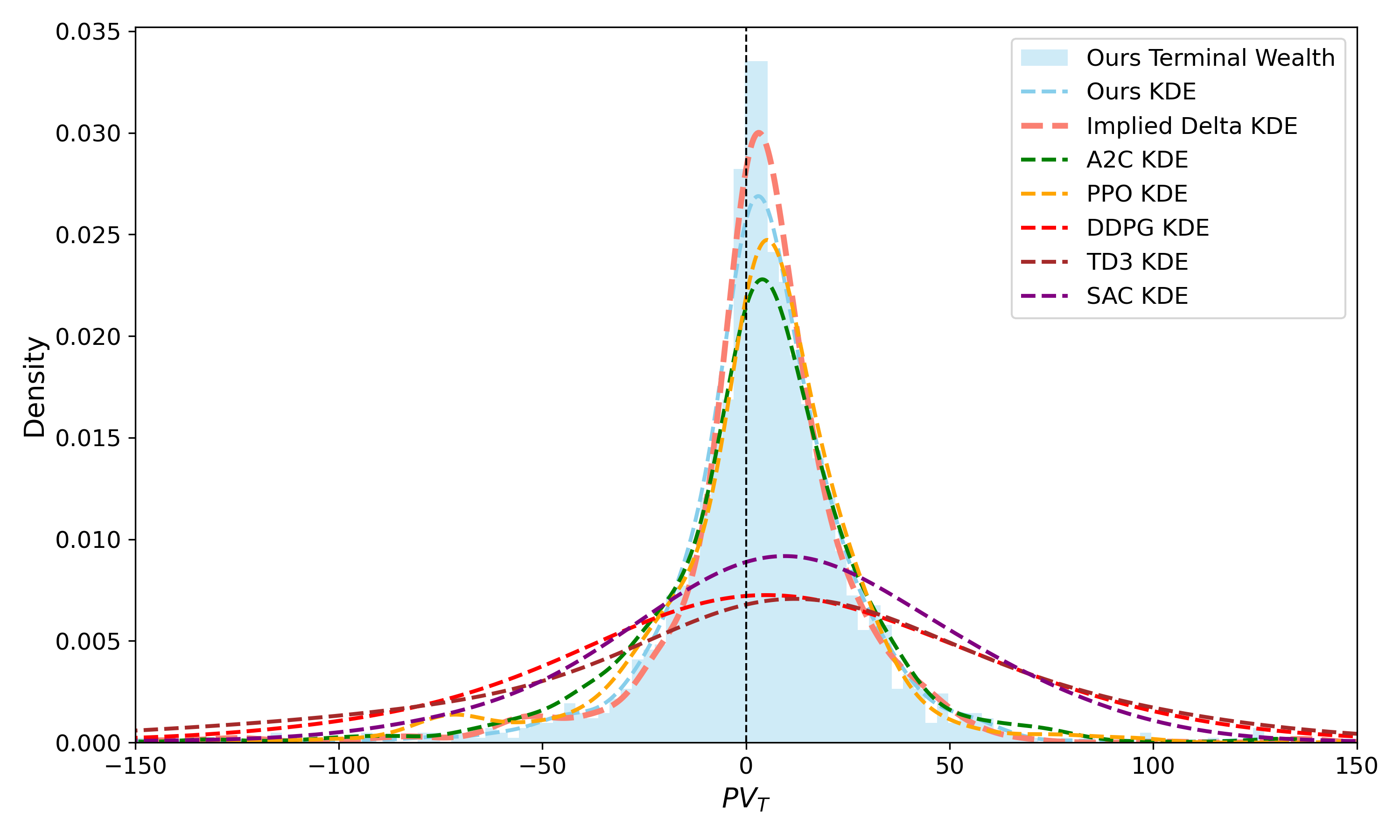}
\caption{Distribution of Portfolio Value at Terminal Time for CSI 300 Index Options under Zero Transaction Costs}
\label{fig:csi300_0tc_pv}
\end{figure}

Figure \ref{fig:csi300_0tc_pv} compares the distributions of $PV_T$ for the Behavior Cloning-Proximal Policy Optimization algorithm, implied Delta, and other baseline reinforcement learning algorithms in the context of CSI 300 index options. For clarity, only the KDE results for the other baseline algorithms are shown. It can be observed that the distribution of $PV_T$ for the BC-PPO algorithm is the most concentrated, indicating the best hedging performance. While the implied Delta, A2C, and PPO have a larger variance in $PV_T$, their control over extreme loss situations is limited. The $PV_T$ distributions for DDPG, TD3, and SAC suggest that their hedging strategies did not converge, resulting in poorer performance.

\begin{figure}[h]
\centering
\includegraphics[width=0.8\textwidth]{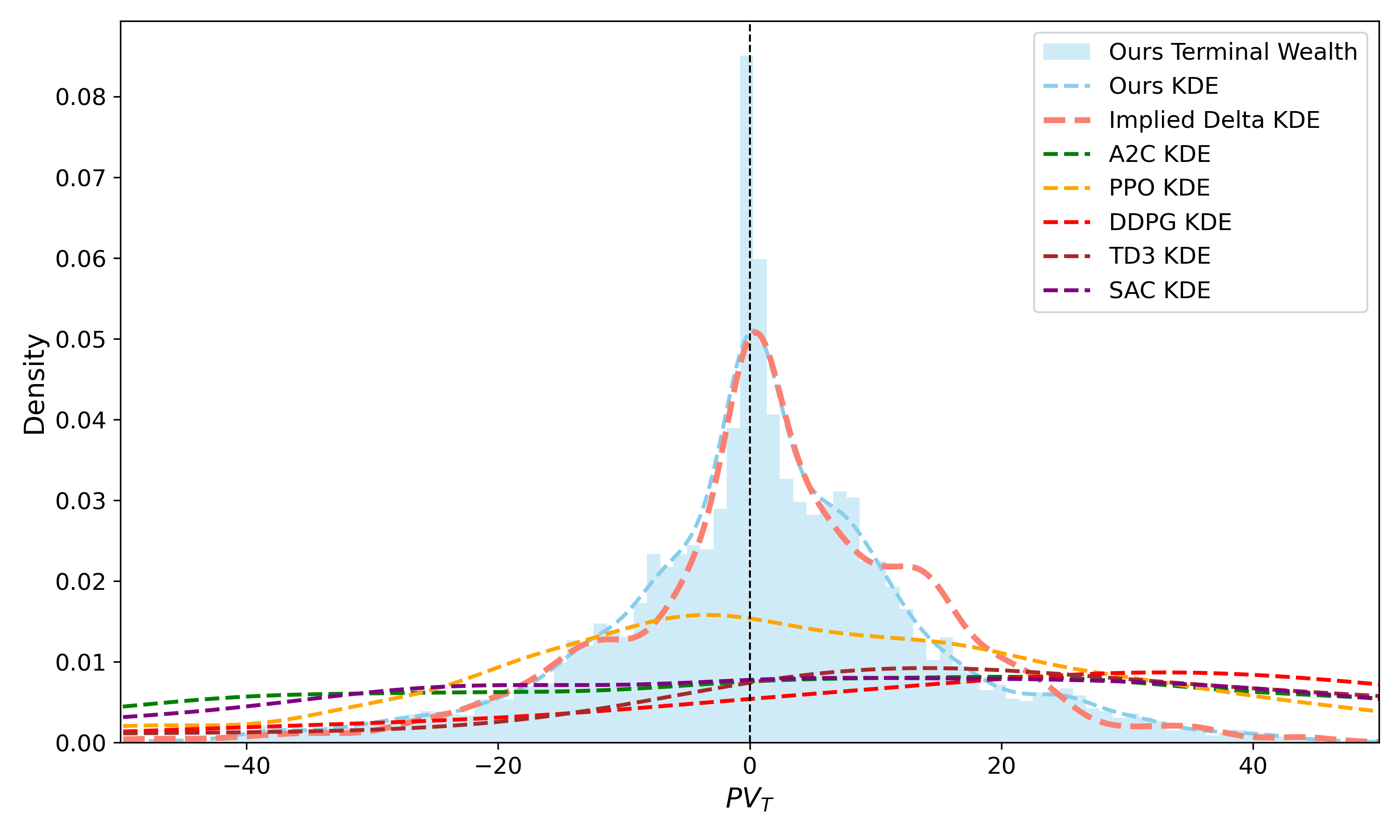}
\caption{Distribution of Portfolio Value at Terminal Time for S\&P 500 Index Options under Zero Transaction Costs}
\label{fig:sp500_0tc_pv}
\end{figure}

Figure \ref{fig:sp500_0tc_pv} compares the distributions of $PV_T$ for the Behavior Cloning-Proximal Policy Optimization algorithm, implied Delta, and other baseline reinforcement learning algorithms in the context of S\&P 500 index options. Compared to implied Delta, the distribution of $PV_T$ for the proposed algorithm is the most concentrated, while other reinforcement learning algorithms, although they may yield positive average returns, show a highly dispersed $PV_T$ distribution. While the proposed algorithm does not significantly outperform A2C and PPO in the Chinese market, it significantly outperforms them in the U.S. stock market, demonstrating its robustness.

\begin{figure}[h]
\centering
\includegraphics[width=0.8\textwidth]{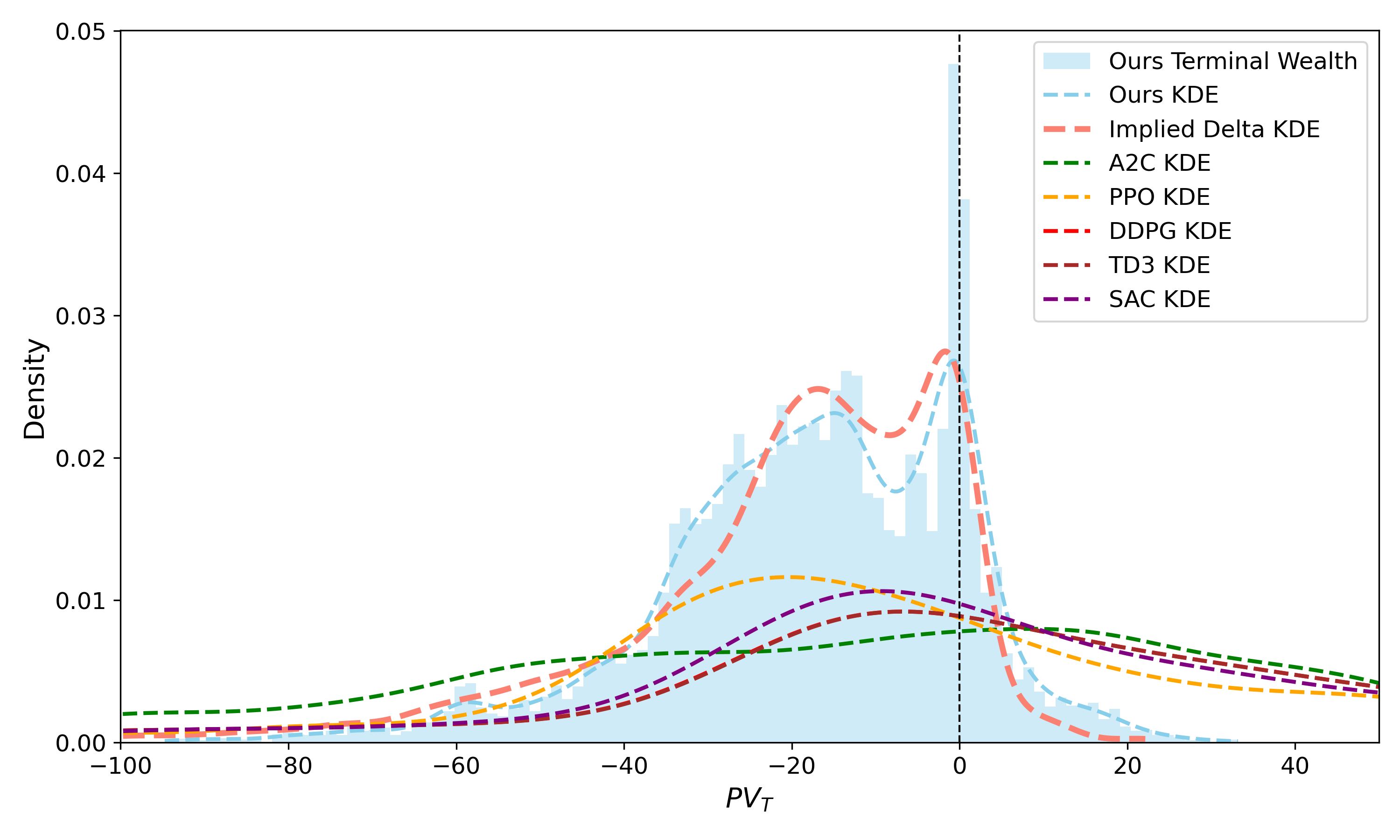}
\caption{Distribution of Portfolio Value at Terminal Time for S\&P 500 Index Options under 0.04\% Transaction Costs}
\label{fig:sp500_4e-3tc_pv}
\end{figure}

Figures \ref{fig:sp500_4e-3tc_pv} compares the distribution of \( PV_T \) for S\&P 500 index options under transaction costs. Since U.S. stock options require more frequent hedging, the mean value of \( PV_T \) for different algorithms is all below zero. Compared to the implied Delta, the proposed algorithm achieves a \( PV_T \) closer to zero and has a higher probability of obtaining positive returns. Other reinforcement learning algorithms struggle more to accurately capture hedging signals under transaction costs.

\subsection{Ablation Study}

The Behavior Cloning (BC) and Recurrent Proximal Policy Optimization (RPPO) algorithm enhances the performance of reinforcement learning by incorporating Behavior Cloning and Recurrent Neural Network (RNN) components. To evaluate their impact, we conducted an ablation study where we compared the proposed algorithm with versions that removed either the Behavior Cloning or RNN components. The study was first conducted under the assumption of zero transaction costs.

\begin{table}[ht]
\centering
\caption{Ablation Study on CSI 300 Index Options under Zero Transaction Costs}
\label{tab:ablation_csi300_zero_tc}
\begin{tabular}{lccc}
\toprule
Model & avg\_r ($\times 10^{-3}$) & avg\_PV & std\_PV \\
\midrule
\textbf{Ours}          & \textbf{-0.5288}  & \textbf{4.2654}  & \textbf{22.5967} \\
Without RNN   & -0.6103  & 3.5567  & 24.4461 \\
Without BC    & -0.6751  & 2.1884  & 25.8894 \\
\bottomrule
\end{tabular}
\end{table}

\begin{table}[ht]
\centering
\caption{Ablation Study on S\&P 500 Index Options under Zero Transaction Costs}
\label{tab:ablation_sp500_zero_tc}
\begin{tabular}{lccc}
\toprule
Model & avg\_r ($\times 10^{-3}$) & avg\_PV & std\_PV \\
\midrule
\textbf{Ours}          & \textbf{-0.1748}  & \textbf{1.2726}  & \textbf{13.1599} \\
Without RNN   & -0.3887  & 1.2116  & 19.6789 \\
Without BC    & -0.9034  & 2.5026  & 29.9529 \\
\bottomrule
\end{tabular}
\end{table}

Tables \ref{tab:ablation_csi300_zero_tc} and \ref{tab:ablation_sp500_zero_tc} show the results of the ablation study on the CSI 300 and S\&P 500 index options under zero transaction costs. After removing either the Behavior Cloning or RNN component, the hedging error increased to varying degrees. From the experimental results, it can be observed that, both in the Chinese and U.S. derivatives markets, the impact of removing Behavior Cloning is significantly greater than that of removing the RNN component. The pre-training provided by Behavior Cloning offers better initial values for the agent, which is more critical than the temporal features perceived by the RNN.

\begin{table}[ht]
\centering
\caption{Ablation Study on CSI 300 Index Options with 0.04\% Transaction Costs}
\label{tab:ablation_csi300_tc}
\begin{tabular}{lccc}
\toprule
Model & avg\_r ($\times 10^{-3}$) & avg\_PV & std\_PV \\
\midrule
\textbf{Ours}          & \textbf{-0.6679}  & \textbf{-5.7988}  & \textbf{25.1856} \\
Without RNN   & -1.7330  & -9.6719  & 40.4908 \\
Without BC    & -4.0608  & -5.0992  & 63.5197 \\
\bottomrule
\end{tabular}
\end{table}

\begin{table}[H]
\centering
\caption{Ablation Study on S\&P 500 Index Options with 0.04\% Transaction Costs}
\label{tab:ablation_sp500_tc}
\begin{tabular}{lccc}
\toprule
Model & avg\_r ($\times 10^{-3}$) & avg\_PV & std\_PV \\
\midrule
\textbf{Ours}          & \textbf{-0.6234}  & \textbf{-17.4727}  & \textbf{17.8364} \\
Without RNN   & -0.7052  & -17.9755  & 19.5481 \\
Without BC    & -1.7065  & -12.6191  & 39.3351 \\
\bottomrule
\end{tabular}
\end{table}

Tables \ref{tab:ablation_csi300_tc} and \ref{tab:ablation_sp500_tc} present the ablation study results on the CSI 300 and S\&P 500 index options under 0.04\% transaction costs. Compared to the removal of the RNN and Behavior Cloning components, the proposed algorithm achieved the optimal hedging error. In the presence of transaction costs, the removal of the Behavior Cloning pre-training component caused a more significant performance degradation. While removing the RNN and Behavior Cloning components may lead to higher returns, the standard deviation of the portfolio value (\(PV_T\)) becomes larger, indicating higher volatility.

\subsection{Impact of Transaction Costs}

Based on the experimental results in Sections 4.3 and 4.4, this paper finds that the presence of transaction costs significantly affects the hedging performance. Therefore, we investigate the impact of different transaction cost rate coefficients, \( c \), in Equation \ref{eq:total_cost} on the hedging performance of the Behavior Cloning - Recurrent Proximal Policy Optimization (BC-RPPO) algorithm, and compare it with the implied Delta.

\begin{figure}[h]
\centering
\includegraphics[width=0.8\textwidth]{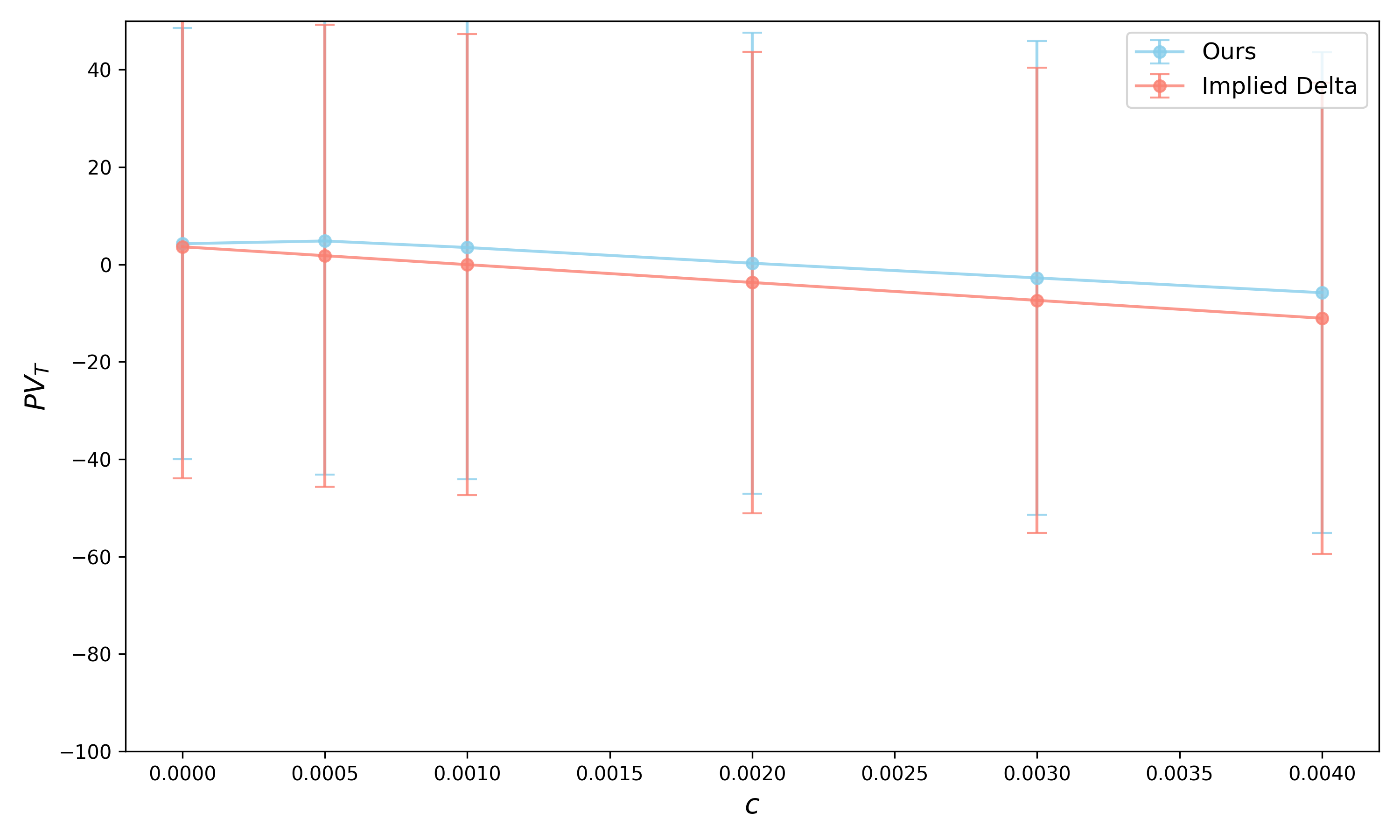}
\caption{Hedging Performance with Different Transaction Cost Rate Coefficients on CSI 300 Index Options}
\label{fig:figure4_13}
\end{figure}

\begin{figure}[h]
\centering
\includegraphics[width=0.8\textwidth]{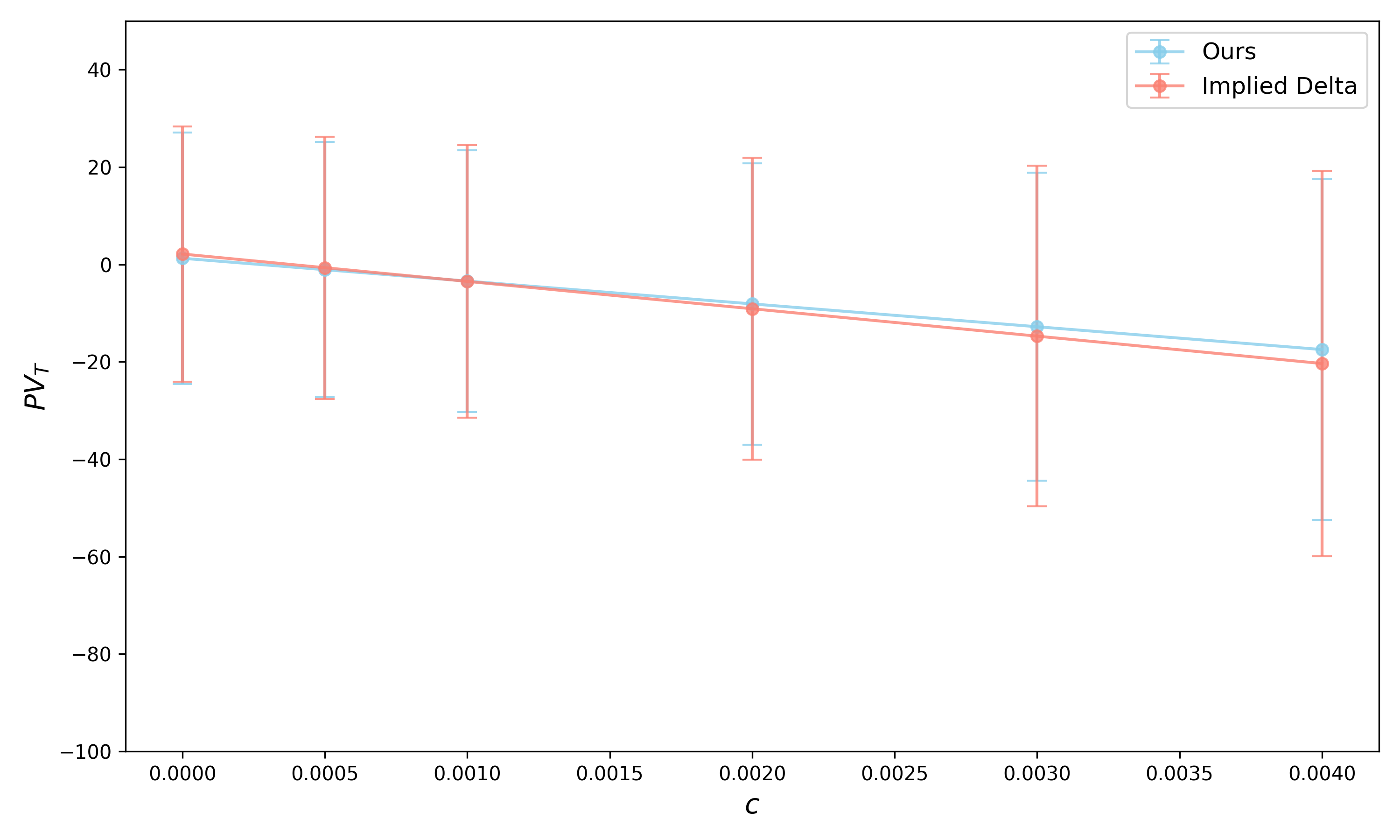}
\caption{Hedging Performance with Different Transaction Cost Rate Coefficients on S\&P 500 Index Options}
\label{fig:figure4_14}
\end{figure}

Figures \ref{fig:figure4_13} and \ref{fig:figure4_14}  display the effects of different transaction cost rate coefficients, \( c \), on the distribution of \( PV_T \) for the CSI 300 and S\&P 500 index options, respectively. The error bars in the figures represent the 95\% confidence interval bounds of \( PV_T \). From the figures, it can be observed that as the transaction cost rate coefficient, \( c \), increases, the difference in the \( PV_T \) distribution between the BC-RPPO algorithm and the implied Delta becomes more pronounced. This indicates that the proposed algorithm remains robust even under high transaction costs.

\subsection{Relationship Between Hedging Performance, Settlement Price, and Contract Expiration Time}

To investigate the relationship between the Behavior Cloning-Cyclic Proximal Policy Optimization (BC-CPPO) algorithm and different settlement prices and contract expiration times, this paper examines the distribution of \( PV_T \) with respect to \( S_t/K \) and \( T-t \) under different transaction costs in the Chinese and U.S. options markets.

\begin{figure}[ht]
\centering
\includegraphics[width=0.8\textwidth]{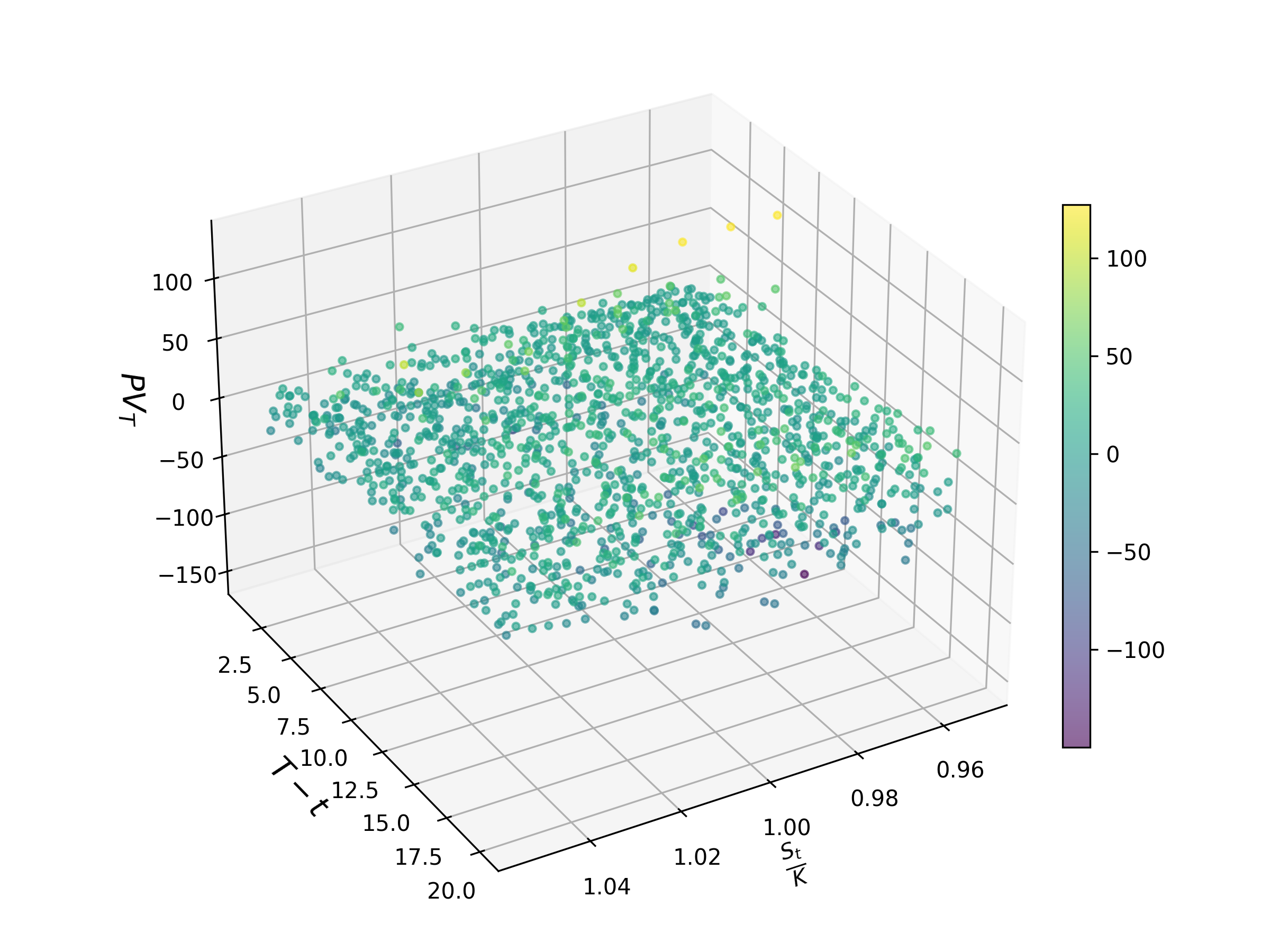}
\caption{Hedging Performance Differences for CSI 300 Index Options Under Zero Transaction Costs}
\label{fig:hedging_csi300_0tc}
\end{figure}

\begin{figure}[ht]
\centering
\includegraphics[width=0.8\textwidth]{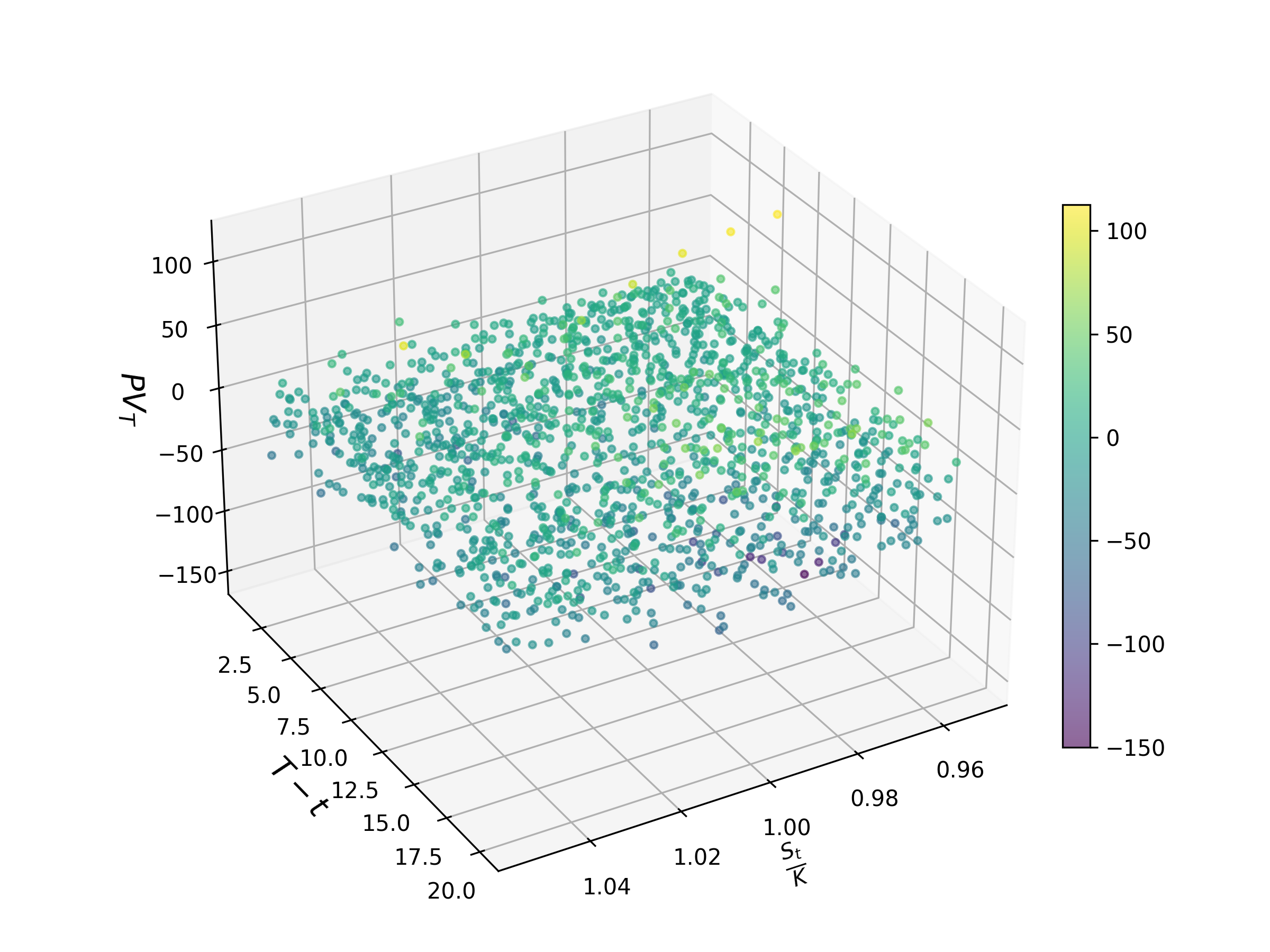}
\caption{Hedging Performance Differences for CSI 300 Index Options Under 0.04\% Transaction Costs}
\label{fig:hedging_csi300_4e-3tc}
\end{figure}

Figures \ref{fig:hedging_csi300_0tc} and \ref{fig:hedging_csi300_4e-3tc} illustrate the distribution of \( PV_T \) with respect to \( S_t/K \) and \( T-t \) for CSI 300 index options under different transaction costs. From the figures, it can be observed that as the contract expiration time \( T-t \) increases, the probability of the total portfolio value \( PV_T \) being negative at the terminal time also increases, indicating that the proposed algorithm is less effective at hedging options with longer expiration times. Additionally, in the hedging task for CSI 300 index options, changes in transaction costs do not significantly alter the relationship between \( PV_T \), \( S_t/K \), and \( T-t \).

\begin{figure}[ht]
\centering
\includegraphics[width=0.8\textwidth]{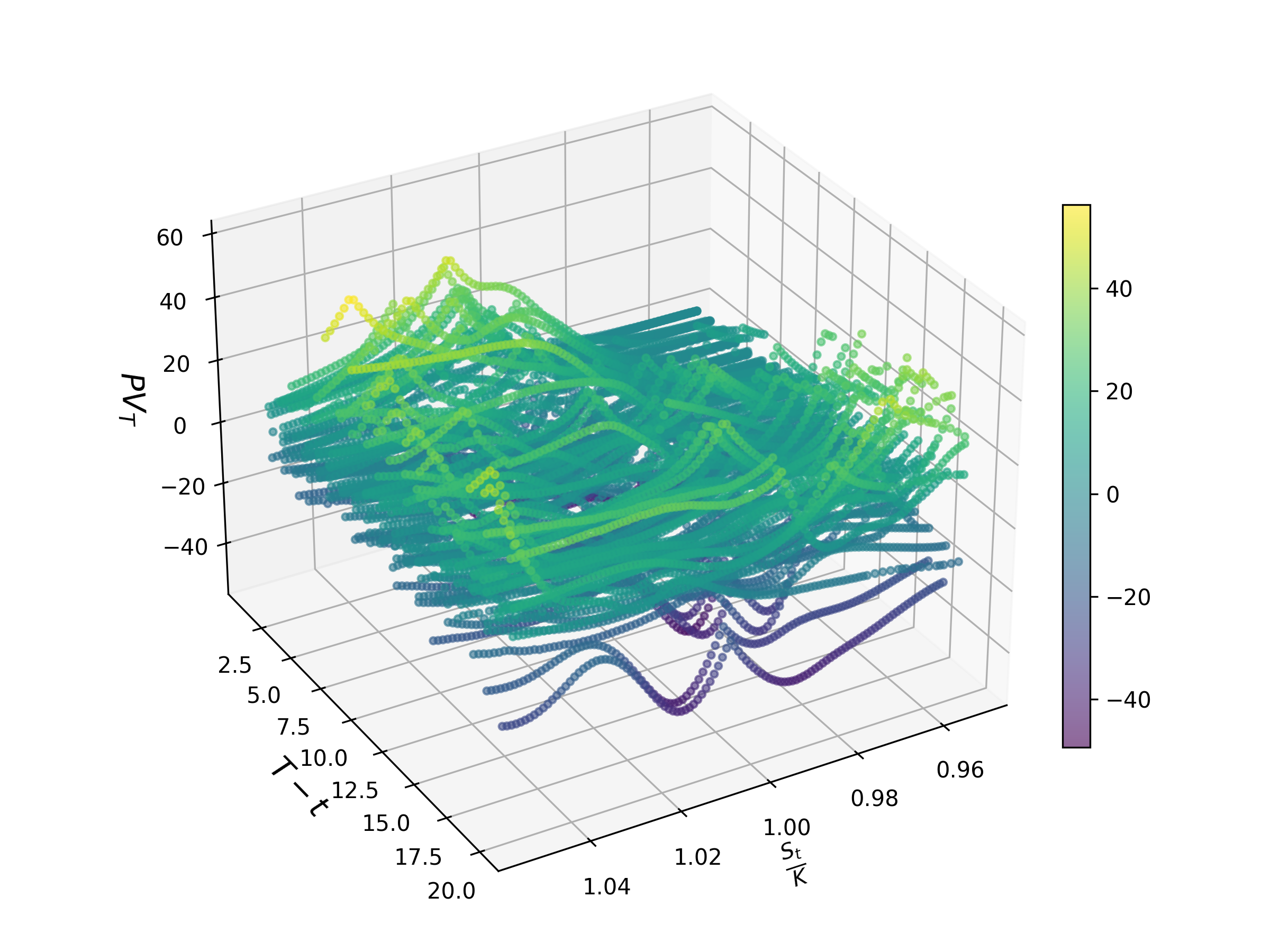}
\caption{Hedging Performance Differences for S\&P 500 Index Options Under Zero Transaction Costs}
\label{fig:hedging_sp500_0tc}
\end{figure}

\begin{figure}[h]
\centering
\includegraphics[width=0.8\textwidth]{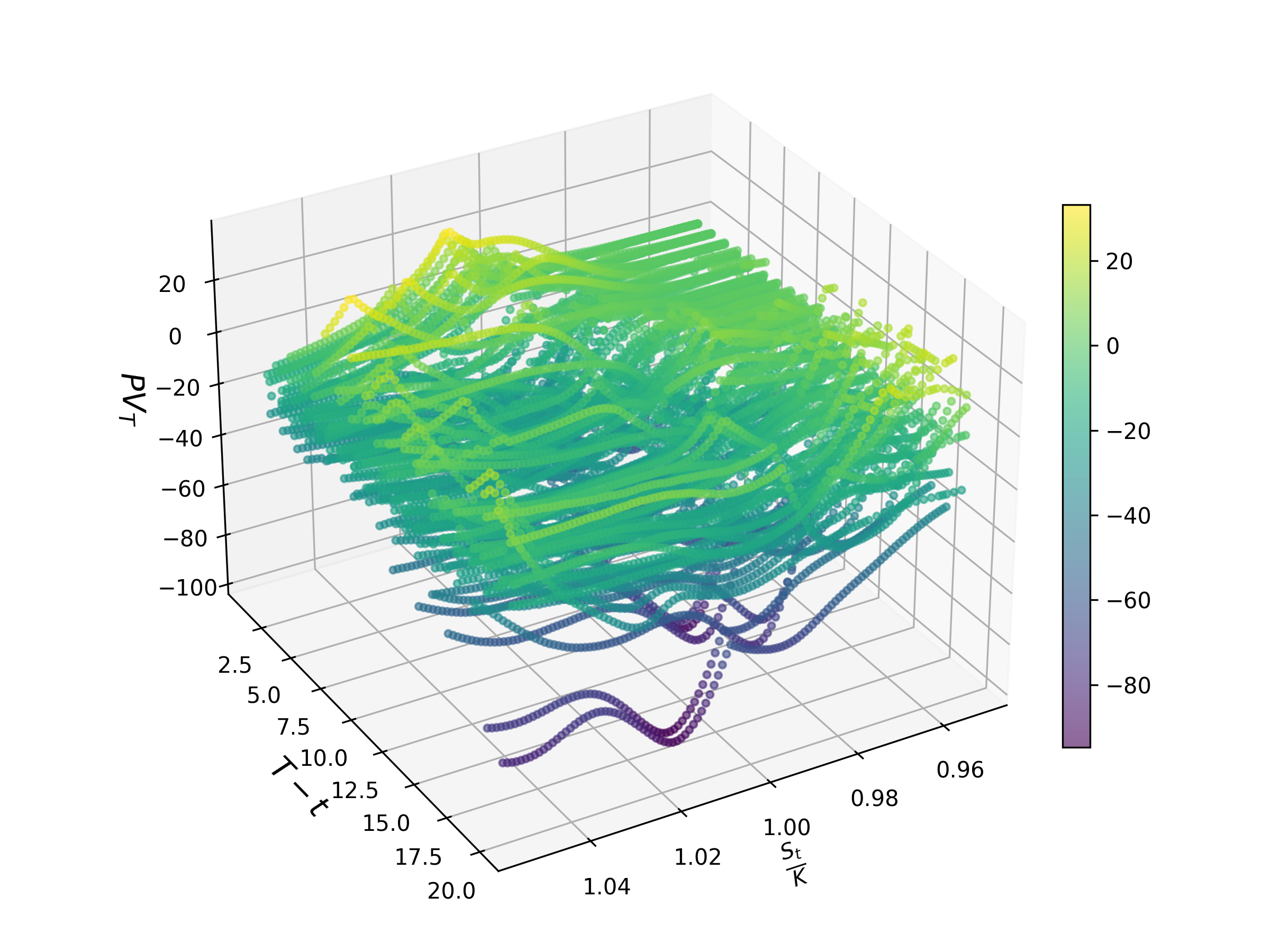}
\caption{Hedging Performance Differences for S\&P 500 Index Options Under 0.04\% Transaction Costs}
\label{fig:hedging_sp500_4e-3tc}
\end{figure}

Figures \ref{fig:hedging_sp500_0tc} and \ref{fig:hedging_sp500_4e-3tc} illustrate the distribution of \( PV_T \) with respect to \( S_t/K \) and \( T-t \) for S\&P 500 index options under different transaction costs. Since the strike price intervals for S\&P 500 index options are set at \$5, while those for CSI 300 index options are set at ¥50, the figures appear denser compared to those of CSI 300 index options. Similarly, it can be observed that the probability of the portfolio incurring large losses increases as the expiration time extends. However, this trend mitigates as \( S_t/K \) approaches 1, reflecting the advantage of the proposed algorithm in hedging at-the-money options.

\section{Conclusion}

This paper addresses the real-world task of derivative hedging, where traditional asset movement equations do not align with real market dynamics, and historical market data used to build offline training environments is often insufficient. We propose a spatiotemporal attention-based probabilistic financial time series prediction Transformer model, which employs low-rank attention to model temporal relationships while simultaneously solving the issue of traditional models being unable to capture the complex nonlinear relationships between stocks. 

Building on this, we provide the theoretical foundation and computational methods for using risk-neutral measures for derivative pricing within deep learning models. The effectiveness of this approach is demonstrated through numerical experiments. Furthermore, by constructing an online training environment based on this model, we integrate derivative hedging tasks with reinforcement learning theory, formulating the task as an agent optimization problem in a sparse reward environment. We introduce the Behavior Cloning - Recurrent Proximal Policy Optimization (BC-RPPO) algorithm, using market implied Delta and maximum likelihood behavior cloning loss for agent pre-training, and fine-tuning the agent using temporal difference error based on \(\lambda\)-returns and generalized advantage estimation.

Additionally, we adopt Recurrent Neural Networks (RNNs) and frame stacking techniques to capture dynamic features in the real market, enhancing the agent's understanding of market information in hedging tasks.

To validate the effectiveness of the proposed algorithm, we utilize real data from the CSI 300 index components and options in China, as well as the S\&P 500 index components and options in the United States. The prediction accuracy and hedging performance of the proposed method are compared with other benchmark methods. Experimental results show that, for prediction tasks, our model achieves the best accuracy, correctly identifying stock time series and complex inter-stock relationships. In hedging tasks, our algorithm delivers the optimal hedging error. Under high transaction cost rate coefficients, our method maintains excellent robustness, significantly outperforming other benchmark algorithms.


\clearpage 

\section*{Acknowledgments}
This work was supported by the National Key Research and Development Program of China (Grant No. 2021ZD0201300), the National Natural Science Foundation of China (Grants No. 12401233 and 12141107).

\bibliographystyle{unsrt}  

\end{document}